\documentclass[12pt]{article}

\usepackage{algorithm, algorithmicx, algpseudocode, hyperref, amssymb, amsfonts, amsmath, amsthm, amsxtra, booktabs, mathrsfs, latexsym, esint, epsfig, epstopdf, graphicx, latexsym, listings, esint, amsxtra, wrapfig, tocvsec2, makecell, multirow, manyfoot, diagbox, color,  xcolor, soul, cases, yfonts, pifont, overpic, enumitem, tikz, tikz-imagelabels, float, textcomp}
\usepackage[mathscr]{eucal}
\usepackage[title]{appendix}%
\usepackage{subfig}

\hypersetup{
     colorlinks=true,
     linkcolor=blue,
     citecolor=green,
     urlcolor=blue
     }

\DeclareMathAlphabet{\mathpzc}{OT1}{pzc}{m}{it}
\allowdisplaybreaks[2]

\def \sn{\,\mbox{sn}}
\def \cn{\,\mbox{cn}}
\def \dn{\,\mbox{dn}}

\title{\boldmath Properties of some elliptic Hill's potentials}

\makeatletter
\renewcommand\@date{{%
  \vspace{0.7cm}%
  \large\centering
  \begin{tabular}{@{}c@{}}
    Wei He\footnote{weihephys@foxmail.com}, \hspace{1 cm}   Peng Su
  \end{tabular}%

  \vspace{0.7cm}

\begin{center}
\em{
  \textsuperscript{}School of Physics and Astronomy, China West Normal University, Nanchong, 637002, China
  }
\end{center}
}}
\makeatother

\begin{document}
\baselineskip=18pt
\maxtocdepth{subsection}
\maketitle

\begin{center}
\textbf{Abstract}
\end{center}
\begin{quotation}
\noindent We study Hill's differential equation with potential expressed by elliptic functions which arises in some problems of physics and mathematics. Analytical method can be applied to study the local properties of the potential in asymptotic regions of the parameter space. The locations of the saddle points of the potential are determined, the locations of turning points can be determined too when they are close to a saddle point. Combined with the quadratic differential associated with the differential equation, these local data give a qualitative explanation for the asymptotic eigensolutions obtained recently. A relevant topic is about the generalisation of Floquet theorem for ODE with doubly-periodic elliptic function coefficient which bears some new features compared to the case of ODE with real valued singly-periodic coefficient. Beyond the local asymptotic regions, global properties of the elliptic potential are studied using numerical method.
\end{quotation}



\newpage

\pagenumbering{arabic}

\section{Elliptic Hill's potential}	

Ordinary differential equations (ODE) are intimately related to applied sciences, for example, a handful of  ODEs  provide the basic mathematical tool to understand problems in physics. In this paper, we study the second order linear ordinary differential equation with a four-term elliptic function coefficient, the coefficient takes the form of a potential given by Eq. (\ref{DTVJacobi}) . It naturally generalises the Lam\'{e}'s differential equation.
The equation is supposed to be compatible with the general theory of differential equation with periodic coefficients developed since the 19th century  \cite{Arscott1964, Magnus-Winkler1966, Eastham1973}; however, it has always been not entirely clear how the Floquet theorem applies for the ODE with elliptic function coefficients. In our previous study, some results in this direction are obtained based on explicit computation of asymptotic spectral solutions \cite{wh1904}. We have computed the spectral solutions in order to study properties of some quantum field theories, therefore, the focus were not on the differential equation itself, certainly there are something more not yet explored.
In this paper, we present some new results about this equation. We aim at providing a more consistent explanation for some basic aspects of  the elliptic function potential that are necessary to understand the spectral solutions obtained recently.

We first introduce some background of Hill's differential equation. In the 19th century, G. W. Hill studied the evolution of the perigee of the moon under the influence of the sun treated as perturbation on the earth-moon system and obtained a second order ODE with periodic coefficients. The problem of solving the differential equation was formulated as solving an algebraic equation of an infinite determinant, in this way Hill obtained perturbative series expansion solution of the differential equation \cite{Hill1877}.
The method introduced by Hill helped astronomers to calculate the trajectory of celestial bodies to very high accuracy,
it influenced the study of  H. Poincar\'{e} on topology inspired by the celestial three-body problem \cite{Gutzwiller1998}.
The differential equation takes the general form
\begin{equation}
\psi^{\,\prime\prime}(x)+[\lambda-u(x)]\psi(x)=0,
\label{DifferentialEq}
\end{equation}
where the coefficient function $u(x)$ is periodic and generally given by trigonometric series,
\begin{equation}
u(x)=\sum_{n=1}^{\infty}\lambda_n\cos(2n x+\phi_n).
\label{HillPotentialTrig}
\end{equation}
When the equation describes a physical problem, the function $u(x)$ is a potential term.
Hill's differential equation is important in quantum mechanics, the solutions of the equation provide the basis to understand band theory of solid physics.
The Schr\"{o}dinger equation for a particle moving in periodic lattice potential is Hill's differential equation,
the spectral solution contains continuous stable regions in the spectral parameter, called energy bands, and continuous unstable regions, called energy gaps. Hill's differential equation is studied using Floquet theorem  \cite{Arscott1964, Magnus-Winkler1966, Eastham1973}.

Elliptic functions are generalisation of trigonometric functions, these meromorphic functions defined in the complex domain have rich analytic properties, they frequently appear in various mathematics and physics problems.
Elliptic functions can be represented in Weierstrass elliptic functions, Jacobian elliptic functions or elliptic theta functions.
In this paper, we use Jacobian form of elliptic functions, there are three related functions, denoted by $\sn\, x\equiv \sn(x,k^2), \cn\, x\equiv \cn(x,k^2)$ and $\dn\, x\equiv \dn(x,k^2)$, where $k^2$ is the elliptic modulus.
We would discuss differential equation (\ref{DifferentialEq}) with elliptic function coefficients, these equations are {\em elliptic Hill's differential equation}, and the potential functions are {\em elliptic Hill's potential}.
The equations of this kind arise in problems of classical mechanics studied in elliptical coordinates,
they have been extensively studied by G. Lam\'e, C. Hermite, J. G. Darboux and E. L. Ince, etc.
In general, the coordinates and parameters of elliptic Hill's differential equation take complex value, the analytic property of the equation and the computation of solutions are not the same as that for the equation with trigonometric potential.
Notably, elliptic Hill's differential equations appear in the study of dynamical systems.
It was discovered in the 1970s that ODEs with elliptic potential are related to elliptic solutions of the Korteweg-de Vries (KdV) hierarchy \cite{novikov1974, lax1975}, the potential is related to meromorphic function on the spectral curve defined by spectral data of the linear differential equation. This connection gives an impetus to study the relation of nonlinear integrable equations and special functions on Riemann surface \cite{Krichever1977}.
There are some particular elliptic potentials associated to spectral curve with finite genus,
in analogy with the case of real potentials with finite number of energy gaps, hence the finite gap potentials.
In the 1990s, A. Treibich and J.-L. Verdier discovered the finite gap potential with four coupling constants \cite{TreibichVerdier1992, TreibichVerdier1994}, it can be expressed using Jacobian elliptic function in the form
\begin{equation}
u(x)=b_0k^2\sn^2x+b_1k^2\frac{\cn^2x}{\dn^2x}+b_2\frac{1}{\sn^2x}+b_3\frac{\dn^2x}{\cn^2x},\label{DTVJacobi}
\end{equation}
or equivalently in the form
\begin{equation}
u(x)=b_0k^2\sn^2x+b_1k^2\sn^2(x+\mathrm{K})+b_2k^2\sn^2(x+i\mathrm{K}^\prime)+b_3k^2\sn^2(x+\mathrm{K}+i\mathrm{K}^\prime),\label{DTVJacobi2}
\end{equation}
where $b_s$ with $s=0, 1, 2, 3$ are the coupling constants, $\mathrm{K}\equiv\mathrm{K}(k^2)$ and $\mathrm{K}^\prime\equiv\mathrm{K}(1-k^2)$ are complete elliptic integrals of the first kind. The potential $u(x)$ has two independent periods $2\mathrm{K}$ and $2i\mathrm{K}^\prime$.
When $b_s/2$ takes the form of triangular number, $b_s=n_s(n_s+1)$, with $n_s\in \mathbb{Z}$, the associated Riemann surface has a finite genus that can be determined by the values of $n_s$. In the study of this paper, the coupling constants $b_s$ are not required to take integer value,
they are general complex parameters. The potential has four singularities in the complex $x$-plane at $x=0$, $\mathrm{K}$, $\mathrm{K}+i\mathrm{K}^{\,\prime}$ and $i\mathrm{K}^{\,\prime}$, module periods.
Hill's differential equation with potential (\ref{DTVJacobi}) is the elliptic form of the Heun's differential equation,
which is the Fuchsian differential equation with four regular singularities.
The differential equation has been studied earlier by Darboux \cite{Darboux1882}, in modern times it arises in some topics of mathematical physics and quantum field theories.
In the theory of isomonodromic deformation, the Heun's differential equation is related to the Painlev\'e VI partial differential equation.
Recently, Hill's differential equation with potential (\ref{DTVJacobi}) appears in the study of quantum field theory,
we have obtained some new results about the asymptotic spectral solutions of differential equation \cite{wh1904}, using computation methods inspired by ideas of quantum field theories  such as the strong-weak duality \cite{SW9407, SW9408} and the relation to integrable quantum mechanical systems \cite{NS0908}.

In more than one century, from time to time the ODE with the potential (\ref{DTVJacobi}) appears in various problems in mathematics,
physics and engineering,
it is a fundamental issue to compute the spectra of this equation to fully understand the phenomenon it describes.
Unlike the few differential equations kept in textbooks, the understanding about ODE with elliptic function coefficient is far from exhaustive, despite many results have been obtained.
The difficulty primarily comes from the fact that the elliptic function potential has second order poles and behaves wildly in the complex plane,
moreover, the equation contains multiple freely-tunable parameters, there is no established method to compute the solution in general.

Based on our earlier study on this problem,
in this paper, we continue to study Hill's differential equation with potential (\ref{DTVJacobi}),
present further analysis on the distribution of the saddle points, turning points and their motion when parameters vary.
The asymptotic spectra was presented in the paper \cite{wh1904} and here we would not repeat the computation;
instead in this paper we focus on some basic aspects of the equation to provide a consistent mathematical background for that computation.
In the following sections, after summing up the spectral results obtained previously, we study the local properties of the elliptic potential in regions  around the saddle points, and we bring turning points to cluster around saddle points or poles by adjusting the eigenvalue.
In these special corners of parameter space, we have the asymptotic spectral solutions obtained recently which give clues about the elliptic generalisation of Floquet theorem.

\section{Summary of results about asymptotic spectral solutions}\label{SummaryAsymptoticSpectralSolutions}

To seek solutions of the equation (\ref{DifferentialEq}), we write the eigenfunction in the exponential form as $\psi(x)=\exp[\int^x v(x^\prime)d x^\prime]$, then the integrand $v(x)$ satisfies the equation $v^2(x)+\partial_xv(x)=u(x)-\lambda$.
There is no general solution of this nonlinear equation, but it is possible to find asymptotic series solutions when an expansion scheme is available. However, there is no obvious reason to assign large or small value to a specific parameter, the study on asymptotic solution is fairly limited as well. In our previous study, a class of asymptotic series solutions are derived since we can identify the correct series expansion parameter  using a connection to some quantum field theories.
A better understanding of the asymptotic spectral solutions obtained there is the main motivation for the study in this paper.
Without repeating the computations carried out in \cite{wh1904}, here we briefly summarise results about the asymptotic spectral solutions as the following three aspects.

\begin{itemize}
\item[(1)] The saddle points of elliptic potential $u(x)$ play a crucial role, in fact, each eigensolution we have obtained is associated with a saddle point, as noted in  \cite{wh1306}. The differential equation has a six-dimensional complex parameter space parameterized by the coupling strength $b_s$ with $s=0, 1, 2, 3$, the elliptic modulus $k$ and the eigenvalue $\lambda$. Because of the complexity of the potential, it is unpractical to derive solutions universally applicable for general parameters. A more practical way is to find solutions that are valid in some specific corners of the parameter space where a large quantity can be identified as the series expansion parameter. These special corners are precisely determined by the saddle points of potential. The potential has six saddle positions, they are classified into three classes according to the local shape; in accordance, there are three families of asymptotic eigensolutions, one weak coupling solution and two strong coupling solutions. We would further elaborate this fact in Sect. \ref{SaddlePointsOfPotential}.
\item[(2)] Around the saddle point $x_*$, the steepness of the potential is characterised by the value of differential $\partial^2_xu(x_*)$, or more precisely by its magnitude. In this paper, the magnitude of a complex quantity refers to its absolute value. From the perspective of quantum theory, the differential equation can be viewed as the Schr\"{o}dinger equation for a quantum particle influenced by the elliptic potential (\ref{DTVJacobi}). When $\partial^2_xu(x_*)$ is large, the particle moving in the steepest direction experiences a very deep potential well, the eigenvalue is approximately that of the infinitely deep square well potential, $\lambda\approx \mu^2$, where $\mu$ is the quantum number and it equals the Floquet index of differential equation. In this case, the eigenvalue is dominated by the kinetic energy, $\lambda\approx\varepsilon_k\gg u(x_{*})\sim b_s$, the eigenvalue $\lambda$, or equivalently the index $\mu$, can be used as the series expansion parameter. There is one series solution of this kind, it is a weak coupling solution because the influence of potential term is very small, and the particles moves almost freely except in the vicinity of boundaries which are singularities of the elliptic function potential. On the other hand, when $\partial^2_xu(x_*)$ is small, the potential is approximately a harmonic potential well with oscillating frequency $2(b_0-b_1)^{1/4}(b_2-b_3)^{1/4}k^{1/2}$, the eigenvalue $\lambda\approx u(x_{*})+4(b_0-b_1)^{1/4}(b_2-b_3)^{1/4}k^{1/2}(\mu+\frac{1}{2})$. In this case, the eigenvalue is dominated by the potential energy, the right expansion parameter is the large frequency, $(b_0-b_1)^{1/4}(b_2-b_3)^{1/4}k^{1/2}\gg \mu$.  There are two series solutions of this kind, they are strong coupling solutions because the oscillating frequency is an effective potential coupling at the saddle position.
\item[(3)] These asymptotic solutions associated with saddle points can be interpreted as Floquet solutions of the differential equation with the elliptic function potential. Compared to the case of trigonometric potential, the peculiarity of the elliptic potential (\ref{DTVJacobi}) is that it has multiple periods, two independent periods $2\mathrm{K}$ and $2i\mathrm{K}^\prime$, the third one $2\mathrm{K}+2i\mathrm{K}^\prime$ also plays a role. It is found that there is a one-to-one correspondence between the three families of spectral solutions and the three periods of elliptic potential. The weak coupling solution is the Floquet solution associated with period $2\mathrm{K}$, the two strong coupling solutions are associated with periods $2i\mathrm{K}^\prime$ and $2\mathrm{K}+2i\mathrm{K}^\prime$, respectively. With this relation, we can compute the eigenvalue $\lambda(\mu)$ and eigenfunction $\psi(\mu,x)$ in a similar way as that in the classical Floquet theorem. We give more details about this point in Sect. \ref{FloquetTheorem4ODEWithEllipticPotential}.
\end{itemize}

Similar features also appear in the simpler case of Mathieu equation and Lam\'{e}'s differential equation, as studied earlier in \cite{wh1108, wh1412, wh1608}.
The differential equation with potential (\ref{DTVJacobi}) is more involved, the results of spectral solution presented in \cite{wh1904} rely much on information input from quantum field theory.
In these ODEs with periodic or elliptic potential, the Floquet index is computed by a contour integral (\ref{FloquetExpIntegral}), this integral formula in quantum mechanics is closely related to the integral formula in Seiberg-Witten gauge theory \cite{SW9407, SW9408}, according to the correspondence revealed in \cite{NS0908}.
In this paper, we try to re-examine the solutions from the view point of differential equation itself.
It is often useful to view the mathematical results of the spectral solution with proper physical background,
as we shall do in later discussion of this paper.
We basically study the shape of the potential (\ref{DTVJacobi}) in the complexified coordinate space,
when zoom in to some special corners the local shape of potential becomes approximately very simple potentials in quantum mechanics, therefore,
the corresponding spectral solutions are qualitatively explained.

\section{Saddle points of the potential}\label{SaddlePointsOfPotential}

\begin{figure}[htbp]
\centering
\subfloat[Global view.]{
\begin{annotationimage}{width=8 cm}{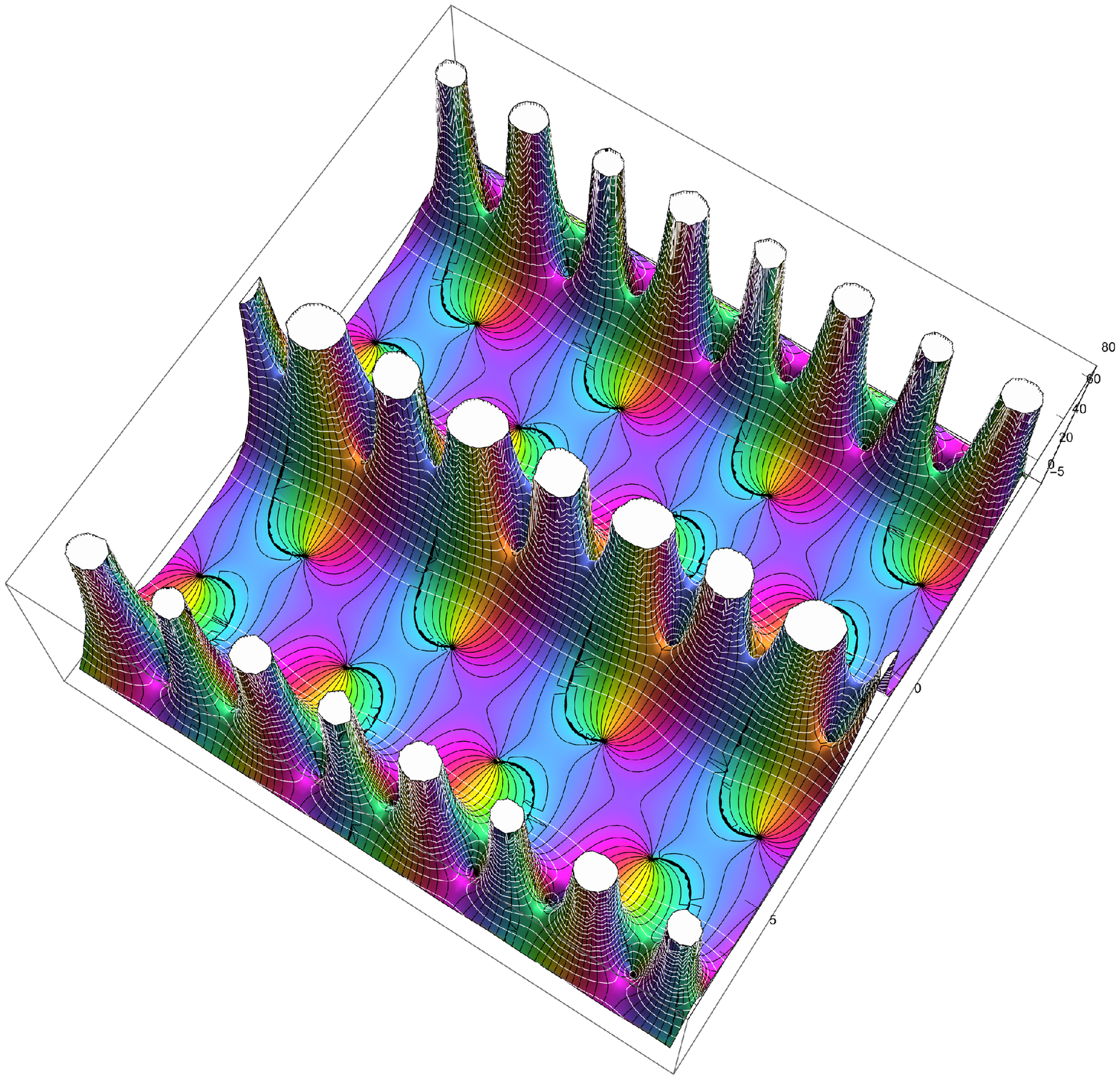}
\node at (1.2, 0.7) {\includegraphics[width=1 cm, height=3 cm]{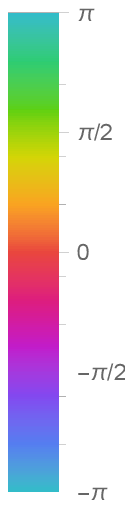}};
\node [black] at (0.1, 0.75) {pole};
\draw[->] (0.15, 0.75) -- (0.285, 0.68); 
\node [black] at (0.0, 0.6) {zero};
\draw[->] (0.05, 0.6) -- (0.268, 0.476); 
\node [black] at (0.2, 0.9) {saddle};
\draw[->] (0.27, 0.89) -- (0.43, 0.79); 
\draw[->] (0.27, 0.89) -- (0.38, 0.679); 
\node [black] at (1.255, 0.35) {saddles: $x_{*1}$, $x_{*2}$, $\cdots$, $x_{*6}$};
\node [black] at (1.247, 0.25) {poles: $0$, $\mathrm{K}$, $i\mathrm{K}^\prime$, $\mathrm{K}+i\mathrm{K}^\prime$};
\node [black] at (1.2, 0.15) {zeros: $x_{1}$, $x_{2}$, $x_{3}$, $x_{4}$};
\end{annotationimage}
\label{LandscapeGlobal}}
\quad
\subfloat[Local view, side-looking.]{
\begin{annotationimage}{width=6.5 cm}{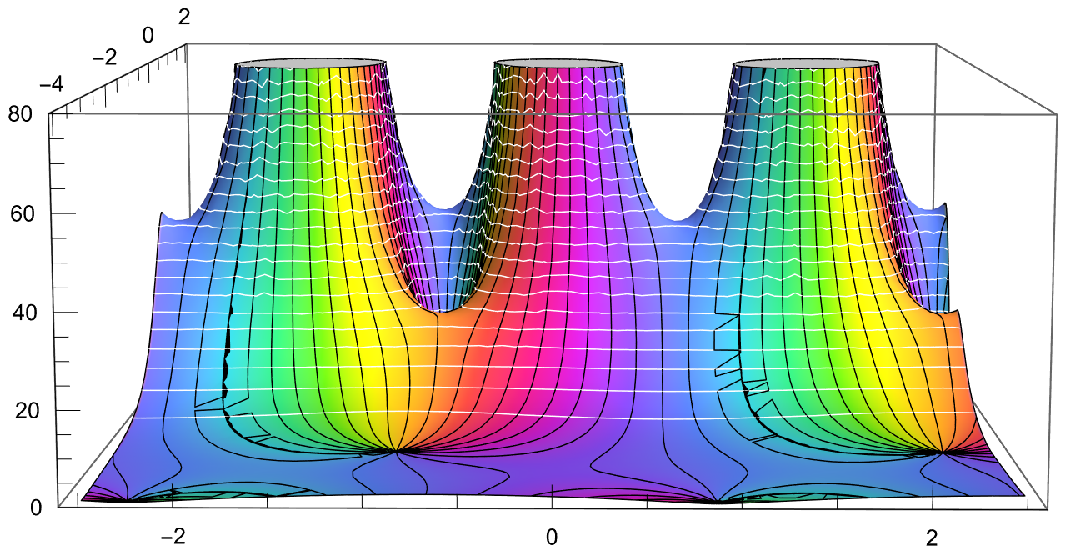}
\node [black] at (-0.03, 0.44) {$u(x_{*3})$};
\draw[->] (0.05, 0.44) -- (0.41, 0.44); 
\node [black] at (-0.03, 0.53) {$u(x_{*4})$};
\draw[->] (0.05, 0.53) -- (0.412, 0.53); 
\node [black] at (-0.03, 0.32) {$u(x_{*5})$};
\draw[->] (0.05, 0.32) -- (0.288, 0.32); 
\end{annotationimage}
\label{LandscapeLocal01}}
\subfloat[Local view, down-looking.]{
\begin{annotationimage}{width=6.5 cm}{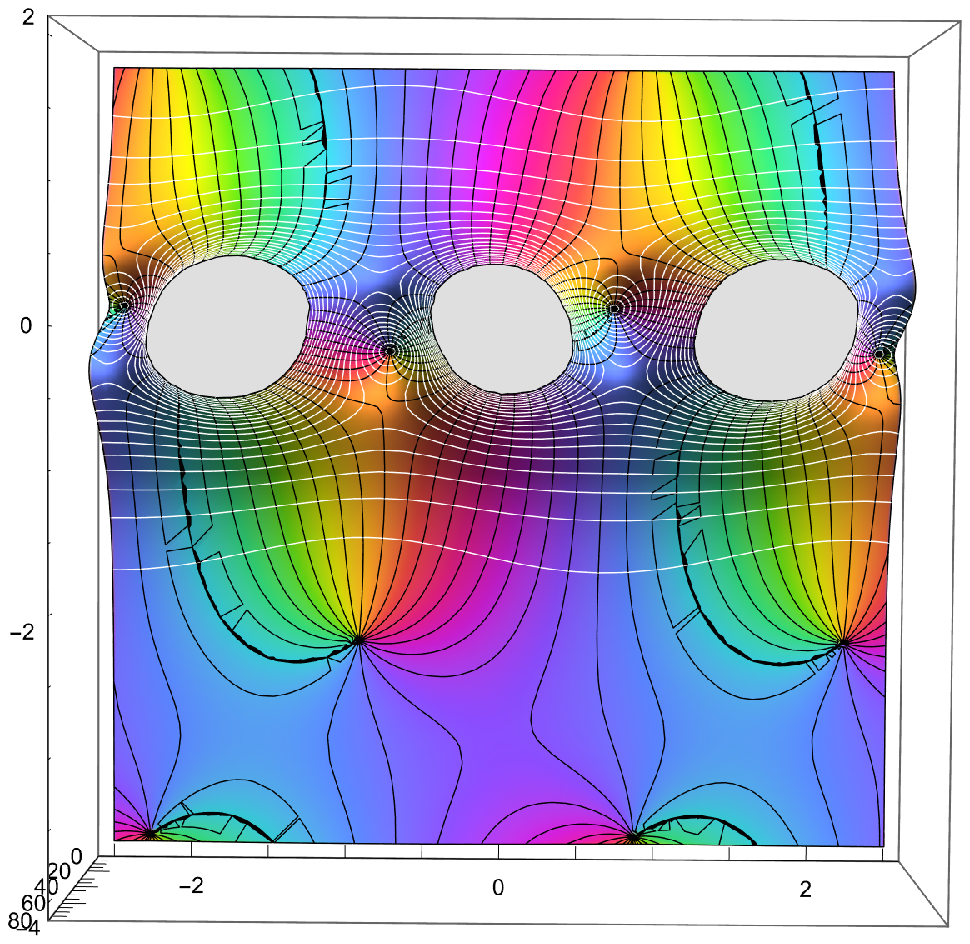}
\node [black] at (0.05, 0.28) {$x_{*5}$};
\draw[->] (0.1, 0.28) -- (0.32,0.28); 
\node [black] at (0.96, 0.28) {$x_{*6}$};
\draw[->] (0.9, 0.28) -- (0.5, 0.28); 
\node [black] at (0.05, 0.36) {$x_{1}$};
\draw[->] (0.1, 0.36) -- (0.391, 0.36); 
\node [black] at (0.05, 0.55) {$x_{*3}$};
\draw[->] (0.1, 0.55) -- (0.41, 0.55); 
\node [black] at (0.415, 1) {$x_{*4}$};
\draw[->] (0.405, 0.95) -- (0.405, 0.645); 
\node [black] at (0.585, 1) {$x_{2}$};
\draw[->] (0.585, 0.95) -- (0.585, 0.625); 
\end{annotationimage}
\label{LandscapeLocal02}}
\vspace{10 pt}
\caption{Stereo view of landscape of the complex potential $u(x)$ over $x$-plane. The height of potential energy surface represents the magnitude $|u(x)|$, with white iso-value contours; the colour shade represents the phase $\mbox{arg}[u(x)]$, with black iso-value contours. 
The high towers are the poles of potential. 
Around the poles are the ``mountain region'' where both $|u(x)|$ and $\mbox{arg}[u(x)]$ change sharply; between them are the ``flatlands''. 
In Fig. \ref{LandscapeGlobal}, one can easily recognize the loci of zeros, where streamlines of the phase come together. 
Iso-value contours are dense around a pole, there are four saddle points at ridges in the mountain region, and two saddle points in the flat region. 
In Fig. \ref{LandscapeLocal01} and Fig. \ref{LandscapeLocal02},
neighbour region around the origin is enlarged. Saddles, poles and zeros are marked up to reflection and period translations.} \label{Landscape}
\end{figure}

The complex potential (\ref{DTVJacobi}) has a complicated landscape in the $x$-plane, labelled by the magnitude $|u(x)|$ and the phase $\mbox{arg}[u(x)]$, analogous to a real valued potential, it has various extremal position, the singularities, the zeros and the saddles, as shown in Fig. \ref{Landscape}. 
All these points play crucial role in the analysis of the potential, the precise location can only be numerically determined for parameters taking generic values, but analytical series solutions can be obtained by carefully adjusting parameters to proper values.

Let us first explain some properties about the saddle points of the potential $u(x)$. The saddle points are determined by the relation
\begin{equation}
\partial_xu(x)|_{x=x_*}=0,
\end{equation}
there are multiple solutions.
To study eigensolutions near a saddle point $x_*$, we expand the potential as a Taylor series,
\begin{equation}
u(x)=u(x_*)+g_2(x-x_*)^2+g_3(x-x_*)^3+\cdots,
\label{HillPotentialSeries1}
\end{equation}
where the coefficients $g_i$ depend on the value of saddle point $x_*$, and in general $g_2\ne 0$, the second term dominates.
The series expansion gives clues about the nature of the asymptotic spectral solutions.

Firstly, at every saddle point the magnitude of the coefficient $g_2$ indicates the form of dispersion relation for the corresponding local solution.
As shown later in this section, the coefficient $g_2$ is large for some saddle points, but small for other saddle points. 
For a large $g_2$, it produces a very deep well around the saddle point. For a particle in a deep well, the energy depends on the quantum number in a quadratic manner, $\lambda\sim \mu^2$.
On the contrary, for a small $g_2$, it produces a very flat potential, approximately a harmonic potential around the saddle point. The energy depends linearly on the quantum number, $\lambda\sim \mu$. 
These preliminary reasonings based on experience of Hermitian quantum mechanics remain valid for the complexified spectral problem, 
the spectral solutions obtained in \cite{wh1904} are indeed in accord with the pattern.

Secondly, the argument of the coefficient $g_s$ determines the steepest trajectories around each saddle point \cite{LiZhou1998}.
The coefficient $g_2$ is complex, its argument is $\alpha_2$; and the argument of $x-x_*$ is $\phi$. Then the second term in the series (\ref{HillPotentialSeries1}) can be written as
\begin{equation}
g_2(x-x_*)^2=|g_2||x-x_*|^2e^{i(\alpha_2+2\phi)}.
\end{equation}
There are two special trajectories crossing the saddle point $x_*$, the one in the direction $\phi=\pi-\alpha_2/2$ is a steepest ascent trajectory, the one in the direction $\phi=\pi/2-\alpha_2/2$ is a steepest descent trajectory.
It is possible to apply the steepest descent method to study the asymptotic series solutions of $\psi(x)$ once it is represented in integral form, similar to the case for Airy function and Bessel function, but it seems more difficult for this equation.
We take a different approach, we would analyse the local form of the elliptic potential around saddle points,
from which properties of the spectral solutions can be inferred.

\begin{figure}[htb]
\vspace{10 pt}
\begin{center}
\includegraphics[width=12 cm]{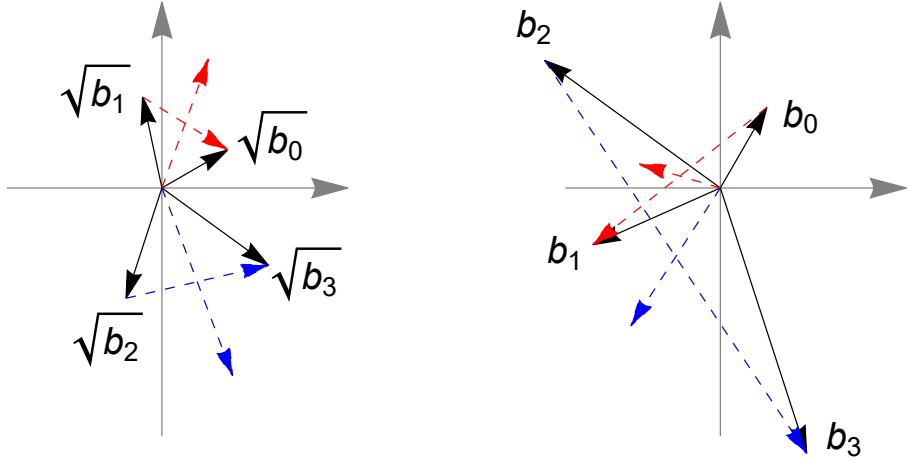}
\end{center}
\vspace{10 pt}
\caption{An example where complex coupling constants $b_s$ take generic values, represented as vectors in the complex plane.
When all magnitudes $|b_s|$ are of the same order, and the angles between $b_0$ and $b_1$, between $b_2$ and $b_3$ are modestly large (say $\gtrsim\frac{\pi}{6}$ and $\lesssim\frac{5\pi}{6}$), then the factors $b_0^{1/2}\pm b_1^{1/2}$ (left chart, red, dashed), $b_2^{1/2}\pm b_3^{1/2}$ (left chart, blue, dashed) all have magnitude of order $\sim\mathcal{O}(\sqrt{b})$, and $b_0\pm b_1$ (right chart, red, dashed), $b_2\pm b_3$ (right chart, blue, dashed) all have magnitude of order $\sim\mathcal{O}(b)$.
The figure is plotted using $b_0^{1/2}=\sqrt{6}\exp\frac{i0.5\pi}{3}$, $b_1^{1/2}=\sqrt{9}\exp\frac{i1.7\pi}{3}$, $b_2^{1/2}=\sqrt{14}\exp\frac{i4.2\pi}{3}$, $b_3^{1/2}=\sqrt{18}\exp\frac{i5.4\pi}{3}$; then for $k$ with value of order $\sim \mathcal{O}(0.01)$, the leading order approximate formulae (\ref{stationarypoints12})-(\ref{coefficientsG3rd}) give fairly accurate results (with the relative error $<10^{-2}$). For non-generic configurations, including the degenerate cases (with nearly parallel or anti-parallel vectors of the same length) and the hierarchical cases (with some vectors much longer than others), the approximate series solutions are not applicable.} \label{GenericValue4Bs}
\end{figure}

Now let us specify to the potential (\ref{DTVJacobi}), its saddle points are determined as follows.
The equation $\partial_x u(x)=0$ can be rewritten as a sextic algebraic equation of $\sn^2x$,
\begin{align}
Q_6(x)=&b_0k^6\sn^{12}x-2b_0(1+k^2)k^4\sn^{10}x \nonumber\\
 &+[b_0(1+4k^2+k^4)-b_1k^{\,\prime\,2}-b_2k^2+b_3k^{\,\prime\,2}k^2]k^2\sn^8x \nonumber\\
 &+2[-b_0(1+k^2)+b_1k^{\,\prime\,2}+b_2(1+k^2)-b_3k^{\,\prime\,2}]k^2\sn^6x \nonumber\\
 &+[b_0k^2-b_1k^{\,\prime\,2}k^2-b_2(1+4k^2+k^4)+b_3k^{\,\prime\,2}]\sn^4x\nonumber\\
 &+2b_2(1+k^2)\sn^2x-b_2=0,
\label{StationaryPointEq}
\end{align}
where $k^{\,\prime}=\sqrt{1-k^2}$ is the complete elliptic module. It has no compact algebraic solution for general parameters $b_s$ and $k^2$. In the case of small elliptic module, $k\ll 1$, there are solutions expanded as series in $k$, they can be divided into three pairs as the following:

{\large \ding{172}} In the neighbourhoods of the first couple saddle points where $\sn^2x\sim\mathcal{O}(\frac{1}{k^2})$, the equation (\ref{StationaryPointEq}) can be approximated by $k^2\sn^8x(b_0k^4\sn^4x-2b_0k^2\sn^2x+b_0-b_1)+\mathcal{O}(k^8)=0$, which contains a second order algebraic equation of $\sn^2x$,
the solution gives the leading order approximation for the saddle position. In our discussion, the leading order result is enough to determine the nature of corresponding spectral solution,
therefore, we present the result as
\begin{equation}
\sn^2x_{*1,2}=\frac{b_0^{1/2}\pm b_1^{1/2}}{b_0^{1/2}}\frac{1}{k^2}+\mathcal{O}(1),
\label{stationarypoints12}
\end{equation}
where plus and minus signs are in accordance with the sub-indices ``1'' and ``2'' labelling the saddle points. With the value, the following quantities associated with $x_{*1,2}$ are determined,
\begin{align}
& u(x_{*1,2})=(b_0^{1/2}\pm b_1^{1/2})^2+\mathcal{O}(k^2),\nonumber\\
& g_2=4(b_0^{1/2}\pm b_1^{1/2})^2+\mathcal{O}(k^2),\nonumber\\
& g_3=\frac{4i(b_0-b_1)(b_0^{1/2}\pm b_1^{1/2})}{b_0^{1/4}(\mp b_1^{1/2})^{1/2}}+\mathcal{O}(k^2),\quad \cdots.
\label{coefficientsG1st}
\end{align}
We consider the case all four coupling constants $b_s$ with $s=0, 1, 2, 3$ and their generic algebraic combinations such as $b_s\pm b_{s^\prime}$, $(b_s^{1/2}\pm b_{s^\prime}^{1/2})^2$, $b_s^{1/2}b_{s^\prime}^{1/2}$, etc, take non-degenerate and non-hierarchical values of the same order of magnitude, denoted by $\mathcal{O}(b)$, see the illustration in Fig.  \ref{GenericValue4Bs}.
In this case, all expansion coefficients $g_m$ listed above have the same order of magnitude, $g_m\sim\mathcal{O}(b)$; moreover, the saddle points $x_{*1}$ and $x_{*2}$ are very close to the poles at $x=\mathrm{K}+i\mathrm{K}^\prime$, up to reflection and translation by periods of the potential. 

{\large \ding{173}} In the neighbourhoods of the second couple saddle points where $\sn^2x\sim\mathcal{O}(1)$, the equation (\ref{StationaryPointEq}) can be approximated by $(b_2-b_3)\sn^4x-2b_2\sn^2x+b_2+\mathcal{O}(k^2)=0$, from which the leading order approximate solution can be calculated. The series solution expands as
\begin{equation}
\sn^2x_{*3,4}=\frac{b_2^{1/2}}{b_2^{1/2}\pm b_3^{1/2}}+\mathcal{O}(k^2),
\label{stationarypoints34}
\end{equation}
the quantities associated with $x_{*3,4}$ are
\begin{eqnarray}
&\quad& u(x_{*3,4})=(b_2^{1/2}\pm b_3^{1/2})^2+\mathcal{O}(k^2),\nonumber\\
&\quad& g_2=4(b_2^{1/2}\pm b_3^{1/2})^2+\mathcal{O}(k^2),\nonumber\\
&\quad& g_3=\frac{4(b_2-b_3)(b_2^{1/2}\pm b_3^{1/2})}{b_2^{1/4}(\pm b_3^{1/2})^{1/2}}+\mathcal{O}(k^2),\quad \cdots.
\label{coefficientsG2nd}
\end{eqnarray}
Likewise, all expansion coefficients $g_m$ have the same order of magnitude, $g_m\sim\mathcal{O}(b)$; the saddle points $x_{*3}$ and $x_{*4}$ are very close to the poles at $x=\mathrm{K}$.

{\large \ding{174}} In the neighbourhoods of the third couple saddle points where $\sn^2x\sim\mathcal{O}(\frac{1}{k})$, the equation (\ref{StationaryPointEq}) can be approximated by $k^2\sn^4x[(b_0-b_1)k^2\sn^4x-(b_2-b_3)]+\mathcal{O}(k^6)=0$, the leading order approximate solution is determined from the nontrivial factor in bracket. The series solution expands as
\begin{equation}
\sn^2x_{*5,6}=\pm\frac{(b_2-b_3)^{1/2}}{(b_0-b_1)^{1/2}}\frac{1}{k}+\mathcal{O}(1),
\label{stationarypoints56}
\end{equation}
the quantities associated with $x_{*5,6}$ are
\begin{eqnarray}
&\quad& u(x_{*5,6})=\pm 2(b_0-b_1)^{1/2}(b_2-b_3)^{1/2}k+\mathcal{O}(k^2),\nonumber\\
&\quad& g_2=\mp 4(b_0-b_1)^{1/2}(b_2-b_3)^{1/2}k+\mathcal{O}(k^2),\nonumber\\
&\quad& g_3=\mp \frac{4i[(b_0-b_1)^2(b_2+b_3)-(b_2-b_3)^2(b_0+b_1)]}{(b_0-b_1)(b_2-b_3)}k^2+\mathcal{O}(k^3),\quad \cdots.
\label{coefficientsG3rd}
\end{eqnarray}
The expansion coefficients have the order of magnitude $g_{2m}\sim\mathcal{O}(bk)$ and $g_{2m+1}\sim\mathcal{O}(bk^2)$.
The saddle points $x_{*5}$ and $x_{*6}$ are away from the poles, sit in the flatland in Fig. \ref{Landscape}.
A particular feature of $x_{*5}$ and $x_{*6}$ is about their distance, from the relation (\ref{stationarypoints56}), one gets
\begin{equation}
\sn^2(x_{*5}+\mathrm{K})=\frac{\cn^2(x_{*5})}{\dn^2(x_{*5})}= -\frac{(b_2-b_3)^{1/2}}{(b_0-b_1)^{1/2}}\frac{1}{k}+\mathcal{O}(1)\approx \sn^2(x_{*6}).
\end{equation}
Therefore, the distance between the pair of saddle points $x_{*5}$ and $x_{*6}$ is approximately $\mathrm{K}$, this fact is used in \cite{wh1904} to compute the corresponding local solutions.

The first two pairs solution given by {\large \ding{172}} and {\large \ding{173}} are expanded as series in $k^2$, the third pair solution given by {\large \ding{174}} is expanded as series in $k$. We here only present the leading order terms in these series expansions, to calculate higher order terms,
we shall expand the solution of $\sn^2x_*$ as a series in $k^2$ or in $k$ and substitute it into the equation (\ref{StationaryPointEq}), the higher order coefficients can be solved order by order. The series expansions of $u(x_*)$ and $g_m$ also can be calculated to arbitrary higher order using  the series expansion of $\sn^2x_*$.
When the parameter $k\ll 1$, for all three pairs of solution the series expansion of $\sn^2x_*$, $u(x_*)$ and $g_m$ are convergent;
by the order of magnitude of the coefficients $g_m$, the radius of convergence for the Taylor expansion of the potential given by (\ref{HillPotentialSeries1}) is $|x-x_*|<1$.

It's now clear how the cases of ``large $\partial^2_xu(x_*)$'' and ``small $\partial^2_xu(x_*)$'' in the previous section are distinguished. Since moderate values for $b_s$ and $k\ll 1$ are assumed, if for {\large \ding{172}} and {\large \ding{173}} $\partial^2_xu(x_*)\sim \mathcal{O}(b)$ is considered to be large, then for {\large \ding{174}} $\partial^2_xu(x_*)\sim \mathcal{O}(bk)$ is small. In Refs \cite{wh1904} and \cite{wh1412, wh1608}, a similar argument is based on ``large potential'' or``small potential'' which are distinguished by the ratio of potential energy and eigenvalue associated with each saddle point. For {\large \ding{172}} and {\large \ding{173}} the potential is small since the ratio $u(x_{*1,2,3,4})/\lambda\sim \mathcal{O}(b/\mu^2)\ll 1$ assuming in this case the index takes large value $\mu\gg b^{1/2}$, while for {\large \ding{174}} the potential is large since the ratio $u(x_{*5})/\delta\sim u(x_{*6})/\widehat{\delta}\sim \mathcal{O}((bk)^{1/2}/\mu)\gg 1$ assuming now the index takes small value $\mu\ll (bk)^{1/2}$, see the eigenvalues given in Sect. \ref{FloquetTheorem4ODEWithEllipticPotential}.

\section{Expansion of the potential at the saddle points}\label{ExpansionPotentialAtSaddlePoints}

To study the local solutions of Hill's differential equation around a saddle point $x_*$, where $x_*$ can be any one of $x_{*r}$ for $r=1, 2,\cdots, 6$,  define the local coordinate $\chi=x-x_*$,
the local potential function $u(\chi)=u(x)-u(x_*)$ and the associated local eigenvalue $\delta=\lambda-u(x_*)$.
Then the equation (\ref{DifferentialEq}) can be rewritten in the form
\begin{equation}
\psi^{\,\prime\prime}(\chi)+[\delta-u(\chi)]\psi(\chi)=0.
\label{DifferentialEq2}
\end{equation}
As the potential (\ref{DTVJacobi}) is an elliptic function, we shall not expand $u(x)-u(x_*)$ as a series in the coordinate $\chi$,
it is more proper to expand in Jacobian elliptic functions, so that the periodicity property of the potential is preserved.
Taking the saddle point $x_{*5}$ as example, to ensure the potential $u(\chi)$ has the same periods $2\mathrm{K}$ and $2i\mathrm{K}^\prime$ as the potential $u(x)$,
the series expansion of $u(x)-u(x_{*5})$ should contain two types of monomials: $\sn^{2m}\chi$ and $\sn^{2m+1}\chi\cn\,\chi\dn\,\chi$ with $m\geqslant 1$
which have exactly periods $2\mathrm{K}$ and $2i\mathrm{K}^\prime$.
The monomials $\sn^{2m}\chi$ are parity even under the reflection $\chi\to -\chi$, and the monomials $\sn^{2m+1}\chi\cn\,\chi\dn\,\chi$ are parity odd,
their series expansions near the origin $\chi=0$ give the even power terms and odd power terms in the series expansion (\ref{HillPotentialSeries1}), respectively.
To compute the series expansion of the local potential $u(\chi)$, one can use the addition formula of elliptic functions to expand every term of the potential (\ref{DTVJacobi}) in the local coordinate.
For example the following formulae are used for $x_*=x_{*5}$,
\begin{equation}
\sn\,x=\sn(x_{*5}+\chi)=\frac{\sn\,x_{*5}\cn\,\chi\dn\,\chi+\sn\,\chi\cn\,x_{*5}\dn\,x_{*5}}{1-k^2\sn^2x_{*5}\sn^2\chi},
\label{snAdditionFormula}
\end{equation}
the right side of the formula can be further expanded as series in $k^2$. There are similar formulae for the functions $\cn\,x$ and $\dn\,x$.
The series expansions are then substituted into the potential (\ref{DTVJacobi}) to compute the difference $u(x)-u(x_{*5})$, in the end we obtain the series expansion of $u(\chi)$ that contains the two types of monomials introduced above.
This is the method used in the reference \cite{wh1904}.

In this paper, we employ another method to compute the local potential function $u(\chi)$ around a saddle point, using the result obtained in Sect. \ref{SaddlePointsOfPotential}.
Let us start from the saddle point $x_{*5}$, the local coordinate is $\chi=x-x_{*5}$.
As explained in Sect. \ref{SaddlePointsOfPotential}, the local potential function $u(\chi)=u(x)-u(x_{*5})$ can be expanded as series in $\chi$,
\begin{equation}
u(\chi)=\sum\limits_{m=1}^{\infty}\left(g_{2m}\chi^{2m}+g_{2m+1}\chi^{2m+1} \right),\label{EllipticHillPotentialExp5}
\end{equation}
where the coefficients $g_m$ are the same as that in the formula (\ref{coefficientsG3rd}) with the upper signs.
At the same time, it also can be expanded as a Taylor series in $\sn\,\chi$,
\begin{equation}
u(\chi)=\sum\limits_{m=1}^{\infty}\left(\lambda_{2m}\sn^{2m}\chi+\lambda_{2m+1}\sn^{2m+1}\chi\cn\,\chi\dn\,\chi \right).
\label{EllipticHillPotentialSN}
\end{equation}
To compute the new coefficients $\lambda_m$, one can expand elliptic functions $\sn\,\chi, \cn\,\chi$ and $\dn\,\chi$ at $\chi=0$,
\begin{subequations}
\begin{align}
& \sn\,\chi=\chi-\frac{1+k^2}{6}\chi^3+\frac{1+14k^2+k^4}{120}\chi^5\cdots,\\
& \cn\,\chi=1-\frac{1}{2}\chi^2+\frac{1+4k^2}{24}\chi^4+\cdots,\\
& \dn\,\chi=1-\frac{k^2}{2}\chi^2+\frac{k^2(4+k^2)}{24}\chi^4+\cdots,
\end{align}
\end{subequations}
substitute the expansions into the series (\ref{EllipticHillPotentialSN}) to convert it into a series in $\chi$.
Comparing the resulting series expansion with that given by (\ref{EllipticHillPotentialExp5}),
the relations between the two groups of coefficients $g_m$ and $\lambda_m$ are obtained,
\begin{subequations}
\begin{align}
& \lambda_2=g_2=-4(b_0-b_1)^{1/2}(b_2-b_3)^{1/2}k+\mathcal{O}(k^2),\\
& \lambda_3=g_3=-\frac{4i[(b_0-b_1)^2(b_2+b_3)-(b_2-b_3)^2(b_0+b_1)]}{(b_0-b_1)(b_2-b_3)}k^2+\mathcal{O}(k^3),\\
& \lambda_4=g_4+\frac{1+k^2}{3}g_2=-\frac{4[(b_0-b_1)^2(b_2+b_3)+(b_2-b_3)^2(b_0+b_1)]}{(b_0-b_1)(b_2-b_3)}k^2+\mathcal{O}(k^3),\\
& \lambda_5=g_5+(1+k^2)g_3=-\frac{6i[(b_0-b_1)^2-(b_2-b_3)^2]}{(b_0-b_1)^{1/2}(b_2-b_3)^{1/2}}k^3+\mathcal{O}(k^4), \quad \cdots.
\end{align}
\end{subequations}
The pattern of magnitude of the coefficients is $\lambda_{2m}\sim\mathcal{O}(b k^{m}), \lambda_{2m+1}\sim\mathcal{O}(b k^{m+1})$.
For $b_s$ and $k$ with finite values, in the limit $m\to\infty$, the limit of higher order term coefficients are $\lambda_m\to 0$ and $\lambda_m^{1/m}\to \sqrt{k}$, then the condition of convergence for the series expansion (\ref{EllipticHillPotentialSN}) is
\begin{equation}
|\sqrt{k}\sn\,\chi| <1.
\label{ConvergenceRegionSN}
\end{equation}
For the case $|k|<1$, the condition implies a roughly belt region in the $x$-plane of width $|i\mathrm{K}^{\,\prime}|$ centered at $x_{*5}$, see Fig. \ref{IntegralContourZ5a}.
In the procedure, the single parameter for series expansion $\lambda_2\approx-4(b_0-b_1)^{1/2}(b_2-b_3)^{1/2}k\sim\mathcal{O}(b k)$ emerges,
it dominates over all other expansion coefficients $\lambda_m$ with $m\geqslant 3$; a price is paid for this benefit, there are infinitely many terms in the locally expanded form of potential.

With explicit expression for the local potential $u(\chi)$ given by (\ref{EllipticHillPotentialSN}), the study could proceed further to calculate the local eigensolution.
In reference \cite{wh1904}, the coefficient $\lambda_2$ is used as the large expansion parameter to calculate the solution $v(\chi)$ of the nonlinear equation $v^2(\chi)+\partial_\chi v(\chi)=u(\chi)-\delta$, it is found that the series expansion of $v(\chi)$ is a Laurent series in $\sn\,\chi$.
Then the Floquet theorem can be applied to compute the eigenvalue. The index $\mu$ is given by the integral of $v(\chi)$ over the period $2i\mathrm{K}^{\,\prime}$,
using the map $\xi=\sn^2\chi$, the integral is converted to the contour integral in the $\xi$-plane so that residue theorem can be applied.
It leads to the large-$\lambda_2$ series expansion $\mu\equiv\mu(\delta)=-\tfrac{1}{2}-\tfrac{1}{2}(-\lambda_2)^{-1/2}\delta+\mathcal{O}(\delta^2)$, which is the inverse dispersion relation.
The eigenfunction is given by the indefinite integral of $v(x)$, as explained in Sect. \ref{SummaryAsymptoticSpectralSolutions}; but it should be noted that the eigenfunction is singular at $\chi=0$.
We do not repeat the calculation in this paper, but some further discussion on the Floquet theorem is presented in Sect. \ref{FloquetTheorem4ODEWithEllipticPotential}.

Near the saddle point $x_{*6}$, one can define another local coordinate $\chi^{\,\prime}=x-x_{*6}$ and expand the potential (\ref{DTVJacobi}) as a series in $\sn\,\chi^{\,\prime}$, the computation would be the same as that for the saddle point $x_{*5}$.
Nevertheless, there is a reason that we should instead expand the local potential around $x_{*6}$ as a series in $\cn\,\chi^{\,\prime}$.
Obtaining the local potential function $u(\chi^{\,\prime})$ is not the ultimate purpose, we want to use $u(\chi^{\,\prime})$ to calculate the series expansion solution of the integrand $v(\chi^{\,\prime})$,
and then integrate it over the period $2\mathrm{K}+2i\mathrm{K}^{\,\prime}$ to derive the corresponding asymptotic eigenvalue.
Using the map $\xi=\sn^2\chi^{\,\prime}$, the integral over  $2\mathrm{K}+2i\mathrm{K}^{\,\prime}$ is converted to the integral over a contour encircling the point $\xi=1$.
But if $v(\chi^{\,\prime})$ is a Laurent series in $\sn\,\chi^{\,\prime}$, it has a singularity at $\chi^{\,\prime}=0$, that is at $\xi=0$ in the $\xi$-coordinate, then the contour integral leads to null result.
The right way is to  use the coordinate defined by $\widehat{\chi}=\chi^{\,\prime}+\mathrm{K}=x-x_{*6}+\mathrm{K}$, because $\cn\,\widehat{\chi}< 1$ around the point $x_{*6}$, then the local potential function can be expanded as a Taylor series in $\cn\,\widehat{\chi}$.
As a result, the integrand $v(\widehat{\chi})$ is a Laurent series in $\cn\,\widehat{\chi}$, it has a singularity at $\widehat{\chi}=\mathrm{K}$ which is mapped to the point $\xi=1$ using the same relation $\xi=\sn^2\widehat{\chi}$.
This point is explained in \cite{wh1904}.

As in the previous case, to maintain the periodicity of elliptic functions,
the expansion of the difference $u(x)-u(x_{*6})$ should contain two types of monomials: $\cn^{2m}\widehat{\chi}$ and $\cn^{2m+1}\widehat{\chi}\sn\,\widehat{\chi}\dn\,\widehat{\chi}$ with $m\geqslant 1$.
To compute the coefficients of the series expansion of $u(\widehat{\chi})=u(x)-u(x_{*6})$, we first expand it as a series in $\chi^{\,\prime}$,
\begin{equation}
u(\widehat{\chi})=\sum\limits_{m=1}^{\infty}\left(\widehat{g}_{2m}\chi^{\,\prime\,2m}+\widehat{g}_{2m+1}\chi^{\,\prime\,2m+1} \right),\label{EllipticHillPotentialExp6}
\end{equation}
where the expansion coefficients $\widehat{g}_m$ are the same as $g_m$ in the formula (\ref{coefficientsG3rd}) with the lower signs.
It also can be expanded as a series in $\cn\widehat{\chi}$,
\begin{equation}
u(\widehat{\chi})=\sum\limits_{m=1}^{\infty}\left(\widehat{\lambda}_{2m}\cn^{2m}\widehat{\chi}+\widehat{\lambda}_{2m+1}\cn^{2m+1}\widehat{\chi}\sn\,\widehat{\chi}\dn\,\widehat{\chi} \right).
\label{EllipticHillPotentialCN}
\end{equation}
Now expand functions $\sn\,\widehat{\chi}, \cn\,\widehat{\chi}$ and $\dn\,\widehat{\chi}$ at the position $\widehat{\chi}=\mathrm{K}$,
which is the same as the position $\chi^{\,\prime}=0$,
\begin{subequations}
\begin{align}
& \sn\,\widehat{\chi}=\frac{\cn\chi^{\,\prime}}{\dn\chi^{\,\prime}}=1-\frac{k^{\,\prime\,2}}{2}\chi^{\,\prime\,2}+\frac{k^{\,\prime\,2}(1-5k^2)}{24}\chi^{\,\prime\,4}+\cdots,\\
& \cn\,\widehat{\chi}=-k^{\,\prime}\frac{\sn\chi^{\,\prime}}{\dn\chi^{\,\prime}}=-k^{\,\prime}\left(\chi^{\,\prime}-\frac{1-2k^2}{6}\chi^{\,\prime\,3}+\frac{1-16k^2+16k^4}{120}\chi^{\,\prime\,5}+\cdots\right),\\
& \dn\,\widehat{\chi}=k^{\,\prime}\frac{1}{\dn\chi^{\,\prime}}=k^{\,\prime}\left(1+\frac{k^2}{2}\chi^{\,\prime\,2}-\frac{k^2(4-5k^2)}{24}\chi^{\,\prime\,4}+\cdots\right),
\end{align}
\end{subequations}
substitute them into the series (\ref{EllipticHillPotentialCN}) to obtain a series in $\chi^{\,\prime}$.
Comparing the resulting series with that given by (\ref{EllipticHillPotentialExp6}) one gets the relation between the two groups of coefficients $\widehat{g}_m$ and $\widehat{\lambda}_m$ as
\begin{subequations}
\begin{align}
& \widehat{\lambda}_2=\frac{1}{k^{\,\prime\,2}}\widehat{g}_2=4(b_0-b_1)^{1/2}(b_2-b_3)^{1/2}k+\mathcal{O}(k^2),\\
& \widehat{\lambda}_3=-\frac{1}{k^{\,\prime\,4}}\widehat{g}_3=\frac{4i[(b_0-b_1)^2(b_2+b_3)-(b_2-b_3)^2(b_0+b_1)]}{(b_0-b_1)(b_2-b_3)}k^2+\mathcal{O}(k^3),\\
& \widehat{\lambda}_4=\frac{1}{k^{\,\prime\,4}}\widehat{g}_4+\frac{1-2k^2}{3k^{\,\prime\,4}}\widehat{g}_2=-\frac{4[(b_0-b_1)^2(b_2+b_3)+(b_2-b_3)^2(b_0+b_1)]}{(b_0-b_1)(b_2-b_3)}k^2+\mathcal{O}(k^3),\\
& \widehat{\lambda}_5=-\frac{1}{k^{\,\prime\,6}}\widehat{g}_5-\frac{1-2k^2}{k^{\,\prime\,6}}\widehat{g}_3=\frac{6i[(b_0-b_1)^2-(b_2-b_3)^2]}{(b_0-b_1)^{1/2}(b_2-b_3)^{1/2}}k^3+\mathcal{O}(k^4), \quad \cdots.
\end{align}
\end{subequations}
The pattern of magnitude of the coefficients is $\widehat{\lambda}_{2m}\sim\mathcal{O}(b k^{m}), \widehat{\lambda}_{2m+1}\sim\mathcal{O}(b k^{m+1})$.
In the limit $m\to\infty$, the limit of the higher order term coefficients are $\widehat{\lambda}_m\to 0$ and $\widehat{\lambda}_m^{1/m}\to \sqrt{k}$,
then the condition of convergence for the series expansion (\ref{EllipticHillPotentialCN}) is
\begin{equation}
|\sqrt{k}\cn\,\widehat{\chi}| <1.
\end{equation}
For $|k|<1$, the condition implies a roughly belt region in the $x$-plane of width $|\mathrm{K}+i\mathrm{K}^{\,\prime}|$ centered at $x_{*6}$.
Similarly, in this case the single parameter for series expansion is $\widehat{\lambda}_{2}\approx4(b_0-b_1)^{1/2}(b_2-b_3)^{1/2}k\sim\mathcal{O}(b k)$, it dominates over all other expansion coefficients $\widehat{\lambda}_m$ with $m\geqslant 3$.
For the local potential function (\ref{EllipticHillPotentialCN}), the solution $v(\widehat{\chi})$ of the nonlinear equation $v^2(\widehat{\chi})+\partial_{\widehat{\chi}}v(\widehat{\chi})=u(\widehat{\chi})-\widehat{\delta}$ is a Laurent series of $\cn\,\widehat{\chi}$.
The index $\mu$ is given by integral of $v(\widehat{\chi})$ over the period $2\mathrm{K}+2i\mathrm{K}^{\,\prime}$,
the resulting inverse dispersion relation is a large-$\widehat{\lambda}_2$ series expansion $\mu\equiv\mu(\widehat{\delta})=-\tfrac{1}{2}-\tfrac{1}{2}\widehat{\lambda}_2^{-1/2}\widehat{\delta}+\mathcal{O}(\delta^2)$, from which the corresponding eigenvalue can be derived.

There remain the saddle points $x_{*r}$ with $r=1,2,3,4$ which share the same nature and can be treated together.
It is indeed possible to expand the potential $u(x)$ around these saddle points, but it turns out that the resulting local form of potential does not help to compute eigensolution. Let us proceed a bit further to see the reason.
To make the local series expansions of potential formally symmetric, we could use the coordinate defined by $\widetilde{\chi}=\chi^{\,\prime\prime}+\mathrm{K}+i\mathrm{K}^\prime=x-x_{*r}+\mathrm{K}+i\mathrm{K}^\prime$, because $\dn\widetilde{\chi}<1$ around the point $x_{*r}$, then the local potential function can be expanded as a Taylor series in $\dn\,\widetilde{\chi}$.
As before, the local potential function $u(\widetilde{\chi})=u(x)-u(x_{*r})$ can be expanded as a series in $\chi^{\,\prime\prime}$,
\begin{equation}
u(\widetilde{\chi})=\sum\limits_{m=1}^{\infty}\left(\widetilde{g}_{2m}\chi^{\,\prime\prime\,2m}+\widetilde{g}_{2m+1}\chi^{\,\prime\prime\,2m+1} \right),\label{EllipticHillPotentialExp1-4}
\end{equation}
where the expansion coefficients $\widetilde{g}_m$ are the same as $g_m$ given in the formulae \eqref{coefficientsG1st} and \eqref{coefficientsG2nd}.
It also can be expanded as a series in $\dn\,\widetilde{\chi}$ as
\begin{equation}
u(\widetilde{\chi})=\sum\limits_{m=1}^{\infty}\left(\widetilde{\lambda}_{2m}\dn^{2m}\widetilde{\chi}+\widetilde{\lambda}_{2m+1}\dn^{2m+1}\widetilde{\chi}\sn\,\widetilde{\chi}\cn\,\widetilde{\chi} \right).
\label{EllipticHillPotentialDN}
\end{equation}
The elliptic functions $\sn\,\widetilde{\chi}, \cn\,\widetilde{\chi}$ and $\dn\,\widetilde{\chi}$ can be expanded at the position  $\widehat{\chi}=\mathrm{K}+i\mathrm{K}^\prime$, that is the same position as $\chi^{\,\prime\prime}=0$, as the following,
\begin{subequations}
\begin{align}
& \sn\,\widetilde{\chi}=\frac{\dn\chi^{\,\prime\prime}}{k\cn\chi^{\,\prime\prime}}=\frac{1}{k}\left(1+\frac{k^{\,\prime\,2}}{2}\chi^{\,\prime\prime\,2}+\frac{k^{\,\prime\,2}(5-k^2)}{24}\chi^{\,\prime\prime\,4}+\cdots\right),\\
& \cn\,\widetilde{\chi}=-\frac{ik^{\,\prime}}{k\cn\chi^{\,\prime\prime}}=-\frac{ik^{\,\prime}}{k}\left(1+\frac{1}{2}\chi^{\,\prime\prime\,2}+\frac{5-4k^2}{24}\chi^{\,\prime\prime\,4}+\cdots\right),\\
& \dn\,\widetilde{\chi}=ik^{\,\prime}\frac{\sn\chi^{\,\prime\prime}}{\cn\chi^{\,\prime\prime}}=ik^{\,\prime}\left(\chi^{\,\prime\prime}+\frac{2-k^2}{6}\chi^{\,\prime\prime\,3}+\frac{16-16k^2+k^4}{120}\chi^{\,\prime\prime\,5}+\cdots\right),
\end{align}
\end{subequations}
substitute them into the series (\ref{EllipticHillPotentialDN}) to convert it to a series in $\chi^{\,\prime\prime}$, and compare the resulting series with that given by (\ref{EllipticHillPotentialExp1-4}), one obtains the relation between the two groups of coefficients $\widetilde{g}_m$ and $\widetilde{\lambda}_m$. In this case, the coefficients $g_m$ have different expressions for each saddle point,
in accordance, the expansion coefficients $\widetilde{\lambda}_m$ also have different expressions. We here take the saddle point $x_{*1}$ as the example, the calculation leads to the following coefficients,
\begin{subequations}
\begin{align}
& \widetilde{\lambda}_2=-\frac{1}{k^{\,\prime\,2}}\widetilde{g}_2=-4(b_0^{1/2}+b_1^{1/2})^2+\mathcal{O}(k^2),\\
& \widetilde{\lambda}_3=-\frac{k^2}{k^{\,\prime\,4}}\widetilde{g}_3=\frac{4(b_0-b_1)(b_0^{1/2}+ b_1^{1/2})}{b_0^{1/4}b_1^{1/4}}k^2+\mathcal{O}(k^4),\\
& \widetilde{\lambda}_4=\frac{1}{k^{\,\prime\,4}}\widetilde{g}_4-\frac{2-k^2}{3k^{\,\prime\,4}}\widetilde{g}_2=\frac{(b_0^{1/2}+b_1^{1/2})^2(5b_0-2b_0^{1/2}b_1^{1/2}+5b_1)}{b_0^{1/2}b_1^{1/2}}+\mathcal{O}(k^2),\\
& \widetilde{\lambda}_5=\frac{1}{k^{\,\prime\,6}}\widetilde{g}_5-\frac{k^2(2-k^2)}{k^{\,\prime\,6}}\widetilde{g}_3=\frac{2(b_0-b_1)(b_0^{1/2}+b_1^{1/2})(3b_0+8b_0^{1/2}b_1^{1/2}+3b_1)}{b_0^{3/4}b_1^{3/4}}k^2+\mathcal{O}(k^4).
\end{align}
\end{subequations}
The pattern of magnitude of the coefficients is $\widetilde{\lambda}_{2m}\sim\mathcal{O}(b), \widetilde{\lambda}_{2m+1}\sim\mathcal{O}(bk^2)$, which is different from the pattern for the coefficients associated with saddle points $x_{*5}$ and $x_{*6}$. In the limit $m\to\infty$, the limit of higher order term expansion coefficients $\widetilde{\lambda}_m$ remain finite, so the condition of convergence for the series expansion (\ref{EllipticHillPotentialDN}) is
\begin{equation}
|\dn\,\widetilde{\chi}|<1.
\end{equation}
Because all expansion coefficients $\widetilde{\lambda}_{2m}$ have the same order of magnitude $\sim\mathcal{O}(b)$,
there is not a single coefficient that can be used to perform series expansion calculation for the eigensolution.
It is clear that the locally expanded form of potential (\ref{EllipticHillPotentialDN}) with infinitely many terms is not conducive to the purpose of deriving asymptotic spectrum.
The studies in \cite{wh1904} and \cite{wh1306} show that the saddle points $x_{*r}$ with $r=1, 2, 3 ,4$ are related to the large eigenvalue spectral solutions with $\lambda\sim \mu^2 \gg b_s$; to compute these solutions, the correct expansion parameter is the eigenvalue $\lambda$, and we shall use the original form (\ref{DTVJacobi}) of the potential.

\section{Turning points of the potential}\label{TurningPointsOfPotential}

The other type of special points associated with the potential are the turning points, they also play an important role in the study of differential equation, it is noticed in \cite{wh1904, wh1306} that asymptotic spectral solutions are always accompanied by turning points coalescing with a saddle point. This phenomenon seems to be connected with the steepest descent method for the differential equation, despite it has not yet been clearly worked out. In this section we study how the distribution of turning points changes as the eigenvalue $\lambda$ varies.

\begin{figure}[htb]
\vspace{10 pt}
\begin{center}
\includegraphics[width=8 cm]{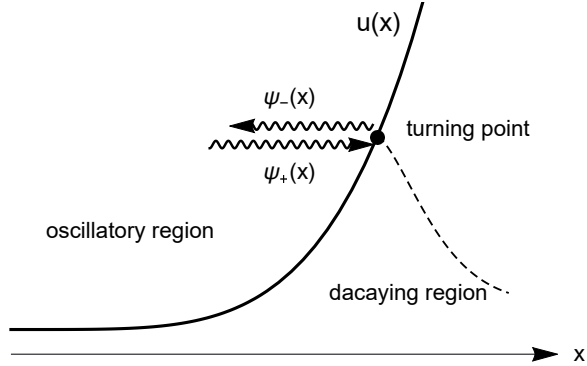}
\end{center}
\vspace{10 pt}
\caption{The right turning point of a real valued potential.} \label{TurningPointRightReal}
\end{figure}

Whenever the right large parameter for series expansion is determined as described in previous section,
one can solve the nonlinear equation for the integrand $v^2(x)+\partial_xv(x)=u(x)-\lambda$ by the Wentzel-Kramers-Brillouin (WKB) series expansion.
The leading order term of the integrand can be written as a differential 1-form,
\begin{equation}
v_0(x)dx= \pm\sqrt{u(x)-\lambda}dx.
\label{DifferentialFormReal}
\end{equation}
The turning points are the zeros of the integrand 1-form, denoted by $x_t$, determined by the equation
\begin{equation}
u(x)-\lambda=0.\label{turningpointsEQ}
\end{equation}
For the real valued potential, the integrand 1-form is complex on one side of a turning point and becomes real on the other side,
in accordance, the eigenfunction $\psi(x)=\exp[\int^x v_0(x^\prime)d x^\prime]$ has oscillatory behaviour on one side and decays on the other side. That means the eigenfunction is truncated at the turning point, the particle is mainly confined in the oscillation region where $u(x)-\lambda<0$.
In the oscillation region, there are two branches of eigenfunction, the right moving and the left moving solutions $\psi_\pm(x)$, as shown in Fig. \ref{TurningPointRightReal}. In a typical situation, near the turning point $x_t$, the potential is approximately linear, $u(x)-\lambda\approx c_{t1}(x-x_t)$, where the coefficient $c_{t1}$ depends on the steepness of the potential at the point $x_t$, the eigenfunction is approximated by the Airy function in this region. The asymptotic behaviours of Airy function on the two sides of the turning point are drastically different,
which precisely describes the transition from the oscillating region to the decaying region.
One can view the turning point as the reflection point for the eigenfunction, where the two branches $\psi_+(x)$ and $\psi_-(x)$ are converted to each other (Fig. \ref{TurningPointRightReal}).

\begin{figure}[hbt]
\begin{minipage}[t]{0.475\linewidth}
\centering
\includegraphics[width=6 cm]{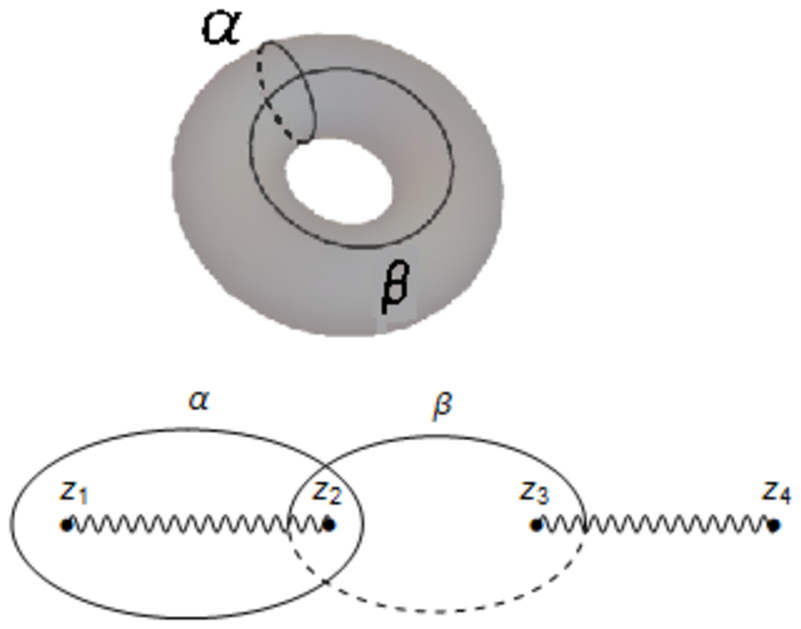}
\vspace{10 pt}
\caption{Turning points, branch cuts and homology bases of the torus.} \label{TurningPointsBranchCuts}
\end{minipage}%
\hspace{0.05\linewidth}
\begin{minipage}[t]{0.475\linewidth}
\centering
\includegraphics[width=5.5 cm]{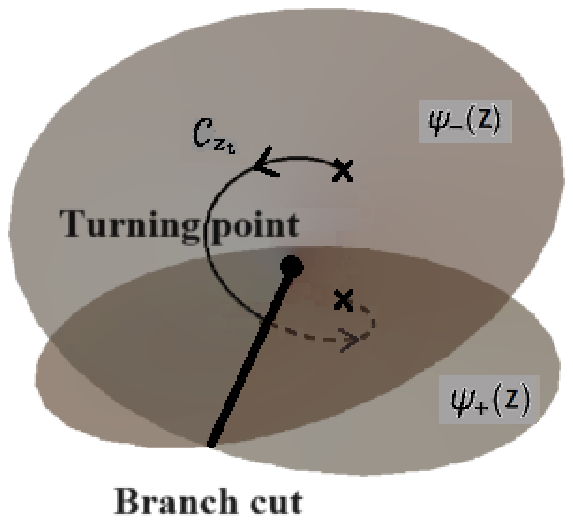}
\vspace{10 pt}
\caption{The path crossing the branch cut from a turning point of the complex potential, the end points on different sheets are images of the same base point $z$.} \label{TurningPointRightComplex}
\end{minipage}
\end{figure}

The properties of turning points has a natural generalisation to the case with complex potential.
Let us specific to the elliptic function potential (\ref{DTVJacobi}), substitute it into the turning point equation (\ref{turningpointsEQ}),
the following quartic equation of $\sn^2x$ is obtained,
\begin{align}
Q_4(x,\lambda)=&\,b_0k^4\sn^{8}x-[(\lambda+b_0-b_1)k^2+(b_0-b_3)k^4]\sn^{6}x \nonumber\\
& +[\lambda+(\lambda+b_0-2b_1+b_2-2b_3)k^2]\sn^{4}x \nonumber\\
& -[\lambda+b_2-b_3-(b_1-b_2)k^2]\sn^{2}x+b_2 \nonumber\\
 =&\,0.
\label{TurningPointEq}
\end{align}
Denote the solutions of the equation by $\sn^2x_t$, $t=1,2,3,4$, which are functions of $b_s$, $k$ and $\lambda$,
then the integrand 1-form is
\begin{equation}
v_0(x)dx=\pm\frac{\sqrt{Q_4(x,\lambda)}}{\sn\,x\cn\,x\dn\,x}dx.
\end{equation}
In the following discussion, it is more convenient to use the variable defined by $z=\sn^{2}x$,
the new coordinate covers the entire complex plane.
The integrand 1-form becomes
\begin{equation}
v_0(z)dz=\pm\frac{\sqrt{b_0\prod_{t=1}^{4}(z-z_t)}}{2z(z-1)(z-k^{-2})}dz,
\label{DifferentialFormComplex}
\end{equation}
where $z_t=\sn^2x_t$.
The integrand has both branch points and poles in the complex $z$-plane, the four branch points are at $z=z_t$ with $t=1,2,3,4$ and the two branch cuts each connects a pair of branch points, as shown in Fig. \ref{TurningPointsBranchCuts}; four simple poles are at $z_p=0, 1, k^{-2},$ and $\infty$.
Therefore, $v_0(z)dz$ is a meromorphic differential 1-form defined on a Riemann surface of genus one which is the covering surface of the branched complex plane,
\begin{equation}
y^2=\prod_{t=1}^{4}(z-z_t),\label{SpectralEllipticCurve}
\end{equation}
with the module parameter defined by the cross ratio,
\begin{equation}
(z_1, z_2, z_3, z_4)\equiv\frac{(z_1-z_2)(z_3-z_4)}{(z_1-z_4)(z_2-z_3)},
\end{equation}
up to $\mathrm{PSL(2,\mathbb{Z})}$ transformations.
This is the same elliptic curve that appears in Seiberg-Witten gauge theory with SU(2) gauge group and $N_f=4$ flavors \cite{SW9408}.

Now we can explain the meaning of turning points for the complex potential of an ordinary differential equation.
The eigenfunction is complex function defined on Riemann surface, on each branch of the surface the eigenfunction is single-valued,
different branches of the eigenfunction are converted to each other by crossing the branch cuts.
Near a turning point $z_t$ there are two eigenfunctions $\psi_{\pm}(z)$ that live on the upper and the lower branch surfaces, the transition between $\psi_+(z)$ and $\psi_-(z)$ is
\begin{equation}
\psi_{\pm}(z+\mathcal{C}_{z_t})=-i\psi_{\mp}(z),
\end{equation}
where $\mathcal{C}_{z_t}$ is a contour crossing the branch cut from $z_t$, see Fig. \ref{TurningPointRightComplex}.
The coefficient $-i$ is inferred from the formal WKB series of eigenfunctions, $\psi_\pm(z)\sim [v_0(z)]^{-1/2}\exp[\int \pm v_0(z)dz+\cdots]$.
The eigenfunctions near the point $z_t$ are approximated by Airy functions, its asymptotic form changes with the argument, this is the Stokes phenomenon. The geometric picture provided by the elliptic curve (\ref{SpectralEllipticCurve}) is closely related to the Floquet theorem for ODE with elliptic potential explained in Sect. \ref{FloquetTheorem4ODEWithEllipticPotential}.
When the eigenfunction translate along the homology circles $\alpha$, $\beta$ or $\alpha+\beta$,
it acquires a phase given by $\exp[\oint v(z)dz]$, this is in fact a manifestation of Floquet theorem for the eigensystem living on the torus.
Our previous study in \cite{wh1904} is largely about how to compute the integrand $v(z)$ when there is a degenerate homology circle, and to identify the correct relation between the contour integral $\oint v(z)dz$ and the Floquet index $\mu$ for the three periods of elliptic function.
A degenerate homology circle always contains one saddle point and a pair turning points closely cluster inside.
The argument can be generalised to the case of higher order complex differential equation and associated surface of higher genus.

\begin{figure}[htb]
\vspace{10 pt}
\begin{center}
\includegraphics[width=6.6cm]{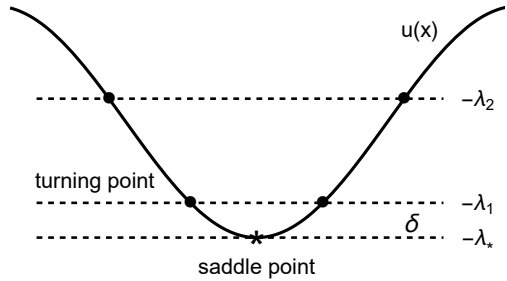}
\end{center}
\vspace{10 pt}
\caption{A pair of turning points approaching the saddle point.} \label{TurningAndStationaryPoints}
\end{figure}

There are algebraic solutions for a quartic polynomial equation, every set of solution $z_t$ determines the shape of the elliptic curve.
For a given potential function $u(x)$, it fixes five parameters $b_s, s=0, 1, 2, 3$ and $k$, the eigenvalue $\lambda$ is tunable,
varying the value of $\lambda$ one can change the shape of the curve.
We are interested in the case when the curve has small cross ratio by  tuning the value of $\lambda$ properly.
There are two ways to achieve this. One can adjust $\lambda$ close to the small value of potential at a saddle point $\lambda_*=u(x_{*r})\sim\mathcal{O}(bk)$, for $r=5, 6$, 
then two adjacent turning points would approach the saddle point $x_{*r}$, the local potential is of the harmonic oscillator type, as shown in Fig. \ref{TurningAndStationaryPoints}.
One can also adjust $\lambda$ a very large value, $\lambda\gg u(x_{*r})\sim\mathcal{O}(b)$, for $r=1, 2, 3, 4$, then two adjacent turning points would asymptotically coalesce with poles at $x=0$ and $x=\mathrm{K}$ respectively, the local potential is of the deep well type,  as shown in Fig. \ref{DeepWellPotential}.
In the following, we compute the series expansion of the solution for the turning point equation when the cross ratio is small,
the relation between saddle points and turning points can be demonstrated explicitly.

\vspace{0.3 cm}
(\uppercase\expandafter{\romannumeral1}) {\large \textbf{Turning points for $\lambda\approx u(x_{*5})$}}
\vspace{0.3 cm}

\begin{figure}[htb]
\vspace{10 pt}
\begin{center}
\includegraphics[width=8.5 cm]{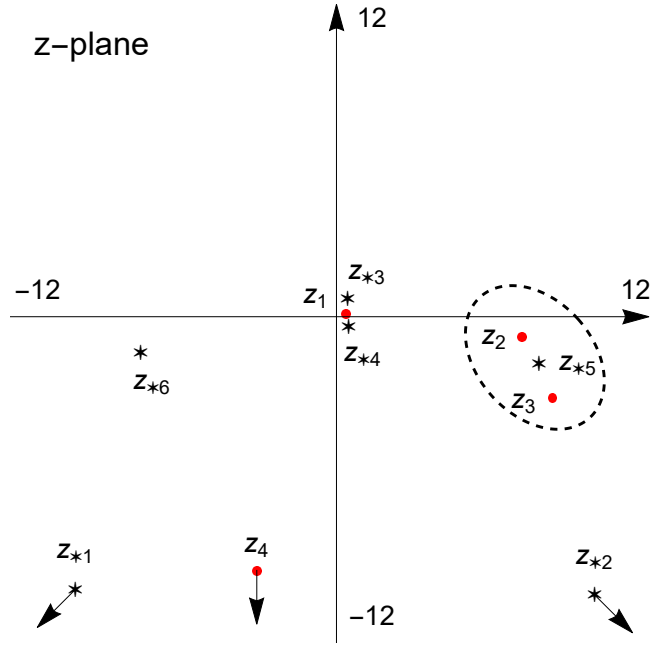}
\end{center}
\vspace{10 pt}
\caption{Distribution of saddle points $z_{*r}$ (black stars) and turning points $z_{t}$ (red dots) in the $z$-plane, for $b_0^{1/2}=\sqrt{6}\exp(i0.5\pi/3)$, $b_1^{1/2}=\sqrt{9}\exp(i1.7\pi/3)$, $b_2^{1/2}=\sqrt{14}\exp(i4.2\pi/3)$, $b_3^{1/2}=\sqrt{18}\exp(i5.4\pi/3)$ and $k=0.2\exp(i\pi/4)$, $\delta=0.1(1+i)$; points $z_{*1}\approx -28.6-13.0i$, $z_{*2}\approx 29.5-34.2i$ and $z_4\approx -2.9-64.3$ are outside the chart. The pair $z_2, z_3$ are adjacent to $z_{*5}$ (enclosed in the dashed oval) while $z_1$ and $z_4$ are scattered elsewhere. To plot the diagram, we have chosen the parameter $k$ with moderate value to avoid too large values of coordinates for these special points, which is not quite appropriate for evaluating the $k$-series expansions in the  previous sections.} \label{DistributionOfPoints5}
\end{figure}

In accordance with the discussion in Sect. \ref{ExpansionPotentialAtSaddlePoints}, we start from the pair of turning points that cluster around the saddle point $x_{*5}$. The elliptic potential (\ref{DTVJacobi}) is very complicated in the whole period parallelogram, but it becomes simpler near a saddle point.
We restrict to the region $|x-x_{*5}|\leqslant \varepsilon$ where $\varepsilon$ is a small quantity.
Correspondingly, the eigenvalue $\lambda$ only differs a small quantity from the value of $u(x_{*5})$, that is
\begin{align}
\lambda=&u(x_{*5})+\delta=2(b_0-b_1)^{1/2}(b_2-b_3)^{1/2}k \nonumber\\
&+\frac{(b_0-b_1-b_2+b_3)(b_1b_2-b_0b_3)}{(b_0-b_1)(b_2-b_3)}k^2+\mathcal{O}(k^3)+\delta,
\label{lambdaNear5th}
\end{align}
where $\delta\ll b k$. Substitute the relation (\ref{lambdaNear5th}) into the equation (\ref{TurningPointEq}),
one gets an algebraic equation about $\sn^2 x$.
The quantity $\delta$ can be used as the small expansion parameter, the equation can be solved order by order.
The leading order approximate equation, denoted as $Q_4^{(0)}(x)=0$, has three solutions, one of them is a degenerate double root exactly coincides with the saddle point $x_{*5}$, therefore $z_{2,3}^{(0)}=\sn^2x_{*5}$; the other two solutions are simple roots. Then turn on a non-zero value for $\delta$, the double roots separate to two simple roots. The solutions of the equation $Q_4(x,\lambda)=0$ are the four turning points as double series expansions in $\delta$ and $k$,
\begin{subequations}
\begin{align}
z_1=&\sum_{m=0}^{\infty}c_{1,m}(b_s,k)\delta^m=\frac{b_2}{b_2-b_3}+\frac{b_2b_3}{(b_2-b_3)^3}\delta+\frac{(b_2+b_3)b_2b_3}{(b_2-b_3)^5}\delta^2+\mathcal{O}\left(\frac{1}{b^3}\delta^3\right),\\
z_2=&\,\sn^2x_{*5}+\sum_{m=1}^{\infty}c_{2,m}(b_s,k)\delta^{m/2}=\sn^2 x_{*5}-\frac{(b_2-b_3)^{1/4}}{(b_0-b_1)^{3/4}k^{3/2}}\delta^{1/2}\nonumber\\
& +\frac{1}{2(b_0-b_1)k^2}\delta -\frac{1}{8(b_0-b_1)^{5/4}(b_2-b_3)^{1/4}k^{5/2}}\delta^{3/2}+\mathcal{O}\left(\frac{1}{b^2k^3}\delta^2\right), \\
z_3=&\,\sn^2x_{*5}+\sum_{m=1}^{\infty}c_{3,m}(b_s,k)\delta^{m/2}=\sn^2x_{*5}+\frac{(b_2-b_3)^{1/4}}{(b_0-b_1)^{3/4}k^{3/2}}\delta^{1/2} \nonumber\\ & +\frac{1}{2(b_0-b_1)k^2}\delta +\frac{1}{8(b_0-b_1)^{5/4}(b_2-b_3)^{1/4}k^{5/2}}\delta^{3/2}+\mathcal{O}\left(\frac{1}{b^2k^3}\delta^2\right), \\
z_4=&\sum_{m=0}^{\infty}c_{4,m}(b_s,k)\delta^m=(1-\frac{b_1}{b_0})\frac{1}{k^2}-\frac{b_1}{b_0(b_0-b_1)k^2}\delta-\frac{b_1}{(b_0-b_1)^3k^2}\delta^2+\mathcal{O}\left(\frac{1}{b^3k^2}\delta^3\right),
\end{align}
\label{TurningPoints5th}%
\end{subequations}
where $c_{t,m}(b_s, k)$ are series in $k$, here we present only the leading order terms in $k$.
In the complex $z$-plane, the four turning points have hierarchical magnitudes as $z_1\sim\mathcal{O}(1)$, $z_2\approx z_3\sim\mathcal{O}(k^{-1})$, $z_4\sim\mathcal{O}(k^{-2})$, so the distance between $z_2$ and $z_3$,
\begin{equation}
z_3-z_2\sim\sqrt{\frac{\delta }{bk^3}}\ll \frac{1}{k},
\label{TuringPoints2and3Distance}
\end{equation}
is much smaller than the distances between $z_2, z_3$ and $z_1$ which are $z_{2,3}-z_1\sim\mathcal{O}(k^{-1})$,
also much smaller than the distances between $z_2, z_3$ and $z_4$ which are $z_{2,3}-z_4\sim\mathcal{O}(k^{-2})$.
This fact indicates $z_2$ and $z_3$ locate closely, and the pair cluster around the saddle point $z_{*5}=\sn^2x_{*5}$.
The distribution of saddle points and turning points is shown in Fig. \ref{DistributionOfPoints5}.
The cross ratio of elliptic curve is
\begin{equation}
(z_3, z_2, z_1, z_4)\approx\frac{2\delta^{1/2}}{(b_0-b_1)^{1/4}(b_2-b_3)^{1/4}k^{1/2}}\sim\mathcal{O}\left(\sqrt{\frac{\delta}{bk}}\right)\ll 1,
\end{equation}
which locates at a singular corner of the moduli space.

\vspace{0.3 cm}
(\uppercase\expandafter{\romannumeral2}) {\large \textbf{Turning points for $\lambda\approx u(x_{*6})$}}
\vspace{0.3 cm}

The turning points associated with the saddle point $x_{*6}$ can be analysed in a similar way.
Now we restrict to the region $|x-x_{*6}|\leqslant \varepsilon$, the corresponding eigenvalue differs a small quantity from the value of $u(x_{*6})$, that is
\begin{align}
\lambda=&u(x_{*6})+\widehat{\delta}=-2(b_0-b_1)^{1/2}(b_2-b_3)^{1/2}k \nonumber\\
&+\frac{(b_0-b_1-b_2+b_3)(b_1b_2-b_0b_3)}{(b_0-b_1)(b_2-b_3)}k^2+\mathcal{O}(k^3)+\widehat{\delta},
\label{lambdaNear6th}
\end{align}
where $\widehat{\delta}\ll b k$. Substitute the relation (\ref{lambdaNear6th}) into the equation (\ref{TurningPointEq}),
one obtains an equation about $\sn^2 x$ and the quantity $\widehat{\delta}$ is used as the small expansion parameter.
The leading order approximate equation $\widehat{Q}_4^{(0)}(x)=0$ has a degenerate double root which coincides with the saddle point $x_{*6}$, that is $\widehat{z}_{2,3}^{(0)}=\sn^2x_{*6}$.
Then turn on a non-zero value for $\widehat{\delta}$, the double root separates to two simple roots. The solutions of the equation $Q_4(x,\lambda)=0$ are the four turning points as double series expansions in $\widehat{\delta}$ and $k$,

\begin{subequations}
\begin{align}
\widehat{z}_1=&\sum_{m=0}^{\infty}\widehat{c}_{1,m}(b_s,k)\widehat{\delta}^{\,m}=\frac{b_2}{b_2-b_3}+\frac{b_2b_3}{(b_2-b_3)^3}\widehat{\delta}+\frac{(b_2+b_3)b_2b_3}{(b_2-b_3)^5}\widehat{\delta}^{\,2}+\mathcal{O}\left(\frac{1}{b^3}\widehat{\delta}^{\,3}\right),\\
\widehat{z}_2=&\,\sn^2x_{*6}+\sum_{m=1}^{\infty}\widehat{c}_{2,m}(b_s,k)\widehat{\delta}^{\,m/2}=\sn^2x_{*6}+\frac{i(b_2-b_3)^{1/4}}{(b_0-b_1)^{3/4}k^{3/2}}\widehat{\delta}^{\,1/2}  \nonumber\\
	& +\frac{1}{2(b_0-b_1)k^2}\widehat{\delta}-\frac{i}{8(b_0-b_1)^{5/4}(b_2-b_3)^{1/4}k^{5/2}}\widehat{\delta}^{\,3/2}+\mathcal{O}\left(\frac{1}{b^2k^3}\widehat{\delta}^{\,2}\right),\\
\widehat{z}_3=&\,\sn^2x_{*6}+\sum_{m=1}^{\infty}\widehat{c}_{3,m}(b_s,k)\widehat{\delta}^{\,m/2}=\sn^2 x_{*6}-\frac{i(b_2-b_3)^{1/4}}{(b_0-b_1)^{3/4}k^{3/2}}\widehat{\delta}^{\,1/2} \nonumber\\
& +\frac{1}{2(b_0-b_1)k^2}\widehat{\delta}+\frac{i}{8(b_0-b_1)^{5/4}(b_2-b_3)^{1/4}k^{5/2}}\widehat{\delta}^{\,3/2}+\mathcal{O}\left(\frac{1}{b^2k^3}\widehat{\delta}^{\,2}\right),\\
\widehat{z}_4=&\sum_{m=0}^{\infty}\widehat{c}_{4,m}(b_s,k)\widehat{\delta}^{\,m}=(1-\frac{b_1}{b_0})\frac{1}{k^2}-\frac{b_1}{b_0(b_0-b_1)k^2}\widehat{\delta}-\frac{b_1}{(b_0-b_1)^3k^2}\widehat{\delta}^{\,2}+\mathcal{O}\left(\frac{1}{b^3k^2}\widehat{\delta}^{\,3}\right),
\end{align}
\label{TurningPoints6th}%
\end{subequations}
where the coefficients $\widehat{c}_{t,m}(b_s, k)$ are series in $k$.
The four turning points have hierarchical magnitudes as $\widehat{z}_1\sim\mathcal{O}(1)$, $\widehat{z}_2\approx \widehat{z}_3\sim\mathcal{O}(k^{-1})$, $\widehat{z}_4\sim\mathcal{O}(k^{-2})$, the distance between $\widehat{z}_2$ and $\widehat{z}_3$ is
\begin{equation}
\widehat{z}_3-\widehat{z}_2\sim \sqrt{\frac{\widehat{\delta}}{bk^3}}\ll \frac{1}{k},
\end{equation}
which is much smaller than the distances between $\widehat{z}_2, \widehat{z}_3$ and $\widehat{z}_1$ which are $\widehat{z}_{2,3}-\widehat{z}_1\sim\mathcal{O}(k^{-1})$, also much smaller than the distances between $\widehat{z}_2, \widehat{z}_3$ and $\widehat{z}_4$ which are $\widehat{z}_{2,3}-\widehat{z}_4\sim\mathcal{O}(k^{-2})$. Therefore, in this case $\widehat{z}_2$ and $\widehat{z}_3$ are close to each other and the pair cluster around the saddle point $z_{*6}=\sn^2x_{*6}$. The cross ratio of elliptic curve is also at a singular corner of the moduli space,
\begin{equation}
(\widehat{z}_2, \widehat{z}_3, \widehat{z}_1, \widehat{z}_4)\approx\frac{-2i\widehat{\delta}^{\,1/2}}{(b_0-b_1)^{1/4}(b_2-b_3)^{1/4}k^{1/2}}\sim\mathcal{O}\left(\sqrt{\frac{\widehat{\delta}}{bk}}\right)\ll 1.
\end{equation}

\begin{figure}[htb]
\vspace{10 pt}
\begin{center}
\includegraphics[width=9 cm]{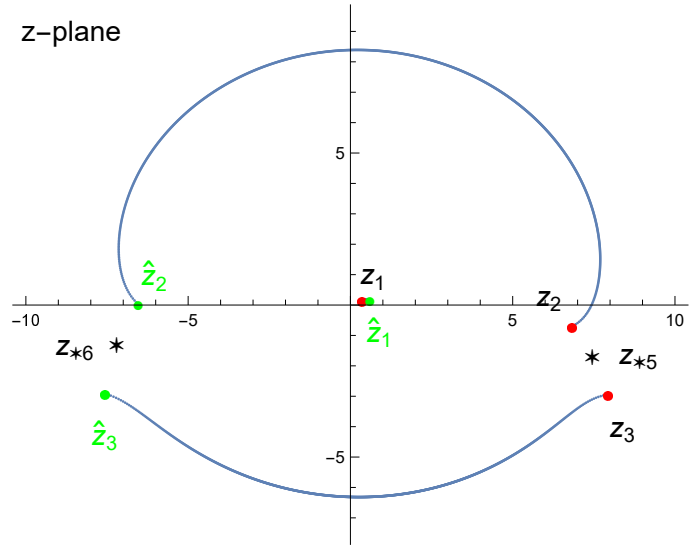}
\end{center}
\vspace{10 pt}
\caption{Trajectories of turning points $z_t$ with $t=1,2,3,4$, the transformation starts from the configuration of $\lambda\approx u(x_{*5})$ (red dots) and ends in the configuration of $\lambda\approx u(x_{*6})$ (green dots). A linear function $\lambda(\varsigma)=u(x_{*5})+\varsigma [u(x_{*6})-u(x_{*5})]+0.1(1+i)$ with $\varsigma\in [0, 1]$ is used to control the moving. Points $z_1$ to $\widehat{z}_1$ are too chose to resolve, and $z_4$ and $\widehat{z}_4$ are outside the chart. Parameters $b_s$ with $s=0, 1, 2, 3$ and $k$ take the same values as that in Fig.  \ref{DistributionOfPoints5}.} \label{TrajectoryOfTurningPoints5To6}
\end{figure}

The solution of turning points are function of eigenvalue, $z_t\equiv z_t(\lambda)$, the turning points move in the $z$-plane as eigenvalue varies.
When the eigenvalue changes its value from $\lambda=u(x_{*5})+\delta$ to $\lambda=u(x_{*6})+\widehat{\delta}$,
the distribution of turning points changes from the one given in (\ref{TurningPoints5th}) to that in (\ref{TurningPoints6th}).
The series expansions of $z_t$ only show the beginning and ending of the moving,
numerical solutions can be used to track the move of individual points.
One can use the following linear function to control the translation of eigenvalue by varying the real parameter $\varsigma$,
\begin{equation}
\lambda(\varsigma)=u(x_{*5})+\varsigma [u(x_{*6})-u(x_{*5})]+\delta.
\end{equation}
In Fig. \ref{TrajectoryOfTurningPoints5To6}, trajectories of turning points are plotted with $\delta=\widehat{\delta}=0.1(1+i)$ when the value of parameter varies in the range  $\varsigma\in [0, 1]$. The intermediate stage of the trajectories corresponds to the portion of parameter space outside the asymptotic regions.
It is clear that, starting from the configuration shown in Fig. \ref{DistributionOfPoints5}, $z_2$ and $z_3$ moves away from the saddle point $z_{*5}$ toward the saddle point $z_{*6}$, and $z_1$ remains almost immobile.

\vspace{0.3 cm}
(\uppercase\expandafter{\romannumeral3}) {\large \textbf{Turning points for $\lambda\gg b_s$}}
\vspace{0.3 cm}

Lastly, we turn to the turning points associated with saddle points $x_{*r}$ for $r=1, 2, 3, 4$.
These four saddle points share the same property, belong to the same class.
As in previous cases, we use $\widetilde{\delta}$ to represent the difference between the eigenvalue $\lambda$ and the value of potential at the saddle point $u(x_{*r})$,
\begin{equation}
\lambda=u(x_{*r})+\widetilde{\delta},
\label{lambdaNear1234th}
\end{equation}
where $u(x_{*r})\sim\mathcal{O}(b)$. According to the explanation in Sect. \ref{SummaryAsymptoticSpectralSolutions},
$\lambda$ is the large parameter used for the series expansion of eigensolution associated to the saddle points $x_{*r}, r=1, 2, 3, 4$,
satisfying  $\lambda\gg b_s$,
therefore, here the difference $\widetilde{\delta}$ must be a large quantity, $\widetilde{\delta}\gg b_s$.
Substitute the relation (\ref{lambdaNear1234th}) into the equation (\ref{TurningPointEq}),
one gets an equation about $\sn^2x$ which can be solved order by order using $\widetilde{\delta}$ as the series expansion parameter. The leading order approximate equation is $Q_4^{(0)}(x)=\widetilde{\delta}\sn^2x(\sn^2x-1)(k^2\sn^2x-1)=0$, there are three simple roots given by $z_1^{(0)}=1$,
$z_2^{(0)}=0$ and $z_3^{(0)}=k^{-2}$. Consider the first order correction by turning on a large but finite value for $\widetilde{\delta}$, the fourth simple root $z_4^{(1)}=\widetilde{\delta}/b_0k^2\gg k^{-2}$ is retrieved. With the leading order solutions, higher order terms of the large-$\widetilde{\delta}$ series expansion can be solved, the roots are double series expansions of $\widetilde{\delta}$ and $k^2$,
\begin{subequations}
\begin{align}
\widetilde{z}_1=&\sum_{m=0}^{\infty}\frac{\widetilde{c}_{1,m}(b_s,k)}{\widetilde{\delta}^{\,m}}=1-\frac{b_3k^{\,\prime\,2}}{\widetilde{\delta}}-\frac{b_3k^{\,\prime\,2}[b_2+b_0k^2+b_3k^2-u(x_{*r})]}{\widetilde{\delta}^{\,2}}+\mathcal{O}\left(\frac{b^3}{\widetilde{\delta}^{\,3}}\right),\\
\widetilde{z}_2=&\sum_{m=1}^{\infty}\frac{\widetilde{c}_{2,m}(b_s,k)}{\widetilde{\delta}^{\,m}}=\frac{b_2}{\widetilde{\delta}}+\frac{b_2[b_3+b_1k^2-u(x_{*r})]}{\widetilde{\delta}^{\,2}} \nonumber\\
&+\frac{b_2[b_2(b_0k^2-b_1k^{\,\prime\,2}k^2+b_3k^{\,\prime\,2})+(b_3+b_1k^2-u(x_{*r}))^2]}{\widetilde{\delta}^{\,3}}+\mathcal{O}\left(\frac{b^4}{\widetilde{\delta}^{\,4}}\right), \\
\widetilde{z}_3=&\sum_{m=0}^{\infty}\frac{\widetilde{c}_{3,m}(b_s,k)}{\widetilde{\delta}^{\,m}}=\frac{1}{k^2}+\frac{b_1k^{\,\prime\,2}}{\widetilde{\delta}k^2}+\frac{b_1k^{\,\prime\,2}[b_0+b_1+b_2k^2-u(x_{*r})]}{\widetilde{\delta}^{\,2}k^2}+\mathcal{O}\left(\frac{b^3}{\widetilde{\delta}^{\,3}k^2}\right), \\
\widetilde{z}_4=&\sum_{m=-1}^{\infty}\frac{\widetilde{c}_{4,m}(b_s,k)}{\widetilde{\delta}^{\,m}}=\frac{\widetilde{\delta}}{b_0k^2}-\frac{b_1+b_3k^2-u(x_{*r})}{b_0k^2}-\frac{b_1k^{\,\prime\,2}+k^2(b_2-b_3k^{\,\prime\,2})}{\widetilde{\delta}k^2}+\mathcal{O}\left(\frac{b^2}{\widetilde{\delta}^{\,2}k^2}\right),
\end{align}
\label{TurningPoints1234th}%
\end{subequations}
where the coefficients $\widetilde{c}_{t,m}(b_s, k)$ are series in $k^2$.
In the above expression of four roots, the expressions of $u(x_{*r})$ for $r=1, 2, 3, 4$ are given in Sect. \ref{SaddlePointsOfPotential}. The four turning points have hierarchical magnitudes as $\widetilde{z}_1\sim 1$, $\widetilde{z}_2\sim 0$, $\widetilde{z}_3\sim k^{-2}$, $\widetilde{z}_4\sim \infty$, the distance between $\widetilde{z}_1$ and $\widetilde{z}_2$ is
\begin{equation}
\widetilde{z}_1-\widetilde{z}_2\sim 1,
\end{equation}
much smaller than the distances between $\widetilde{z}_1, \widetilde{z}_2$ and $\widetilde{z}_3$ which are $\widetilde{z}_{1,2}-\widetilde{z}_3\sim\mathcal{O}(k^{-2})$, also much smaller the distances between $\widetilde{z}_1, \widetilde{z}_2$ and $\widetilde{z}_4$ which are $\widetilde{z}_{1,2}-\widetilde{z}_4\sim \infty$. The cross ratio of elliptic curve is
\begin{equation}
(\widetilde{z}_2, \widetilde{z}_1, \widetilde{z}_3, \widetilde{z}_4)\approx\frac{k^2}{1-k^2}\sim\mathcal{O}(k^2)\ll 1,
\end{equation}
which locates at another singular corner of the moduli space. In the series expansions of $\widetilde{z}_1, \widetilde{z}_2, \widetilde{z}_3, \widetilde{z}_4$ given in (\ref{TurningPoints1234th}), $u(x_{*r})$ appears in higher order terms of the $\widetilde{\delta}$-series expansion,
so the turning points universally approach the poles at $z=0, 1, k^{-2}$ and $\infty$, respectively, and the cross ratios obtained for $\lambda$ in (\ref{lambdaNear1234th}) with different $x_{*r}$ only differ by quantities of order $\mathcal{O}(\widetilde{\delta}^{\,-2})$.

\begin{figure}[htb]
\vspace{10 pt}
\begin{center}
\begin{overpic}[width=11 cm]{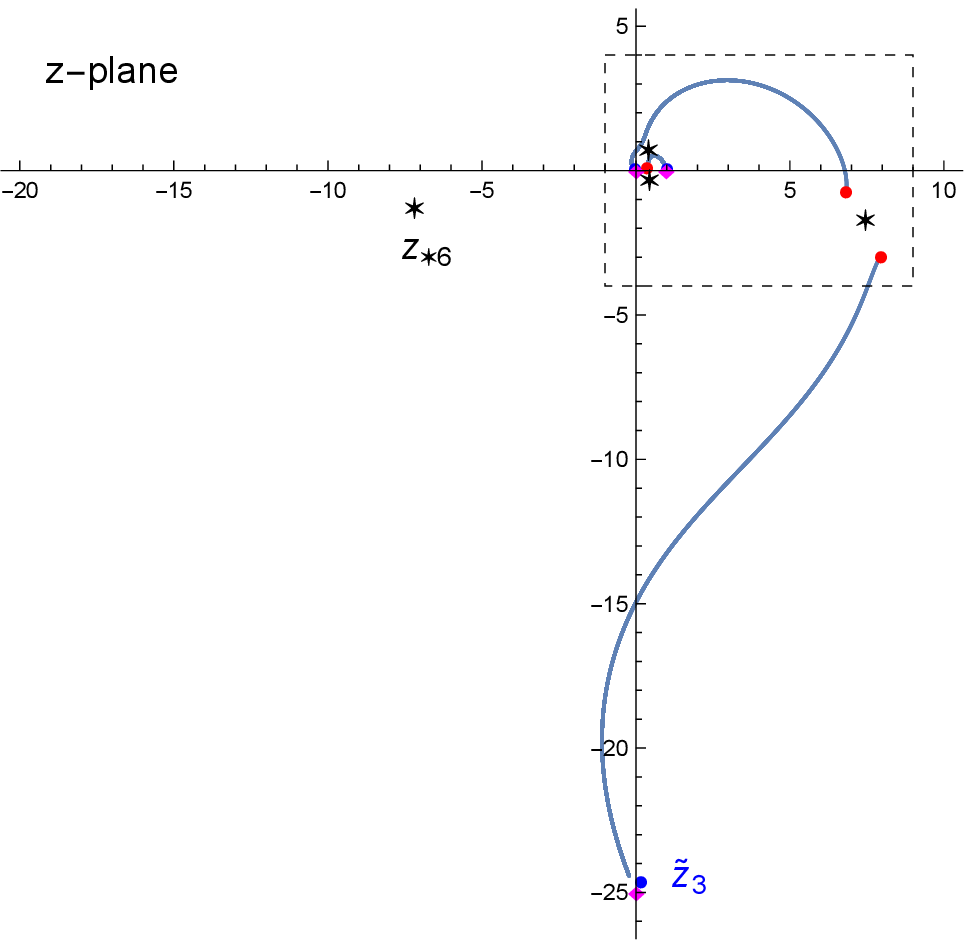}
\put(-10, 5){\includegraphics[width=7 cm]{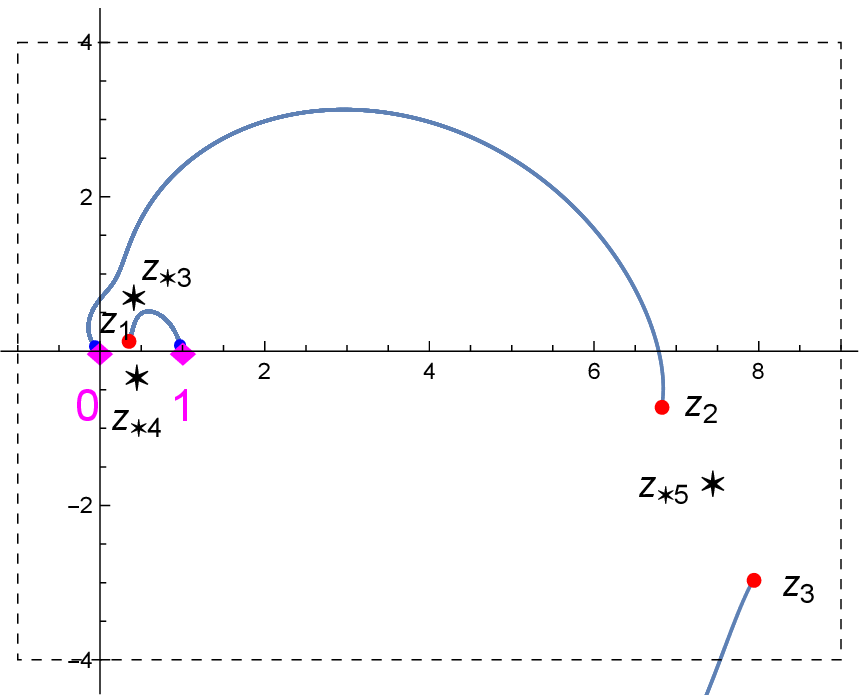}}
\end{overpic}
\end{center}
\vspace{10 pt}
\caption{Trajectories of turning points $z_t$ with $t=1, 2, 3, 4$, the transformation starts from the configuration of $\lambda\approx u(x_{*5})\sim\mathcal{O}(bk)$ (red dots) and ends in the configuration of $\lambda\approx\widetilde{\delta}\gg b_s$ (blue dots) where $z_t$ are very close to the poles at $z=0, 1, k^{-2}$ and $\infty$ (pink diamonds). The trajectory of $z_4$ to $\widetilde{z}_4$ is outside the chart. The region around the origin in the dashed rectangle is enlarged to show more clearly the details. The function  $\lambda(\varsigma)=u(x_{*5})+0.1(1+i)+1000\{1-\exp[0.05\varsigma/(\varsigma-1)]\}$ with $\varsigma\in [0, 0.90]$ is used to plot.
Parameters $b_s$ with $s=0, 1, 2, 3$ and $k$ take the same values as that in Fig. \ref{DistributionOfPoints5}.} \label{TrajectoryOfTurningPoints5To1}\end{figure}

\begin{figure}[hbt]
\centering
\subfloat[A real valued confining potential.]{\includegraphics[width=0.35\linewidth]{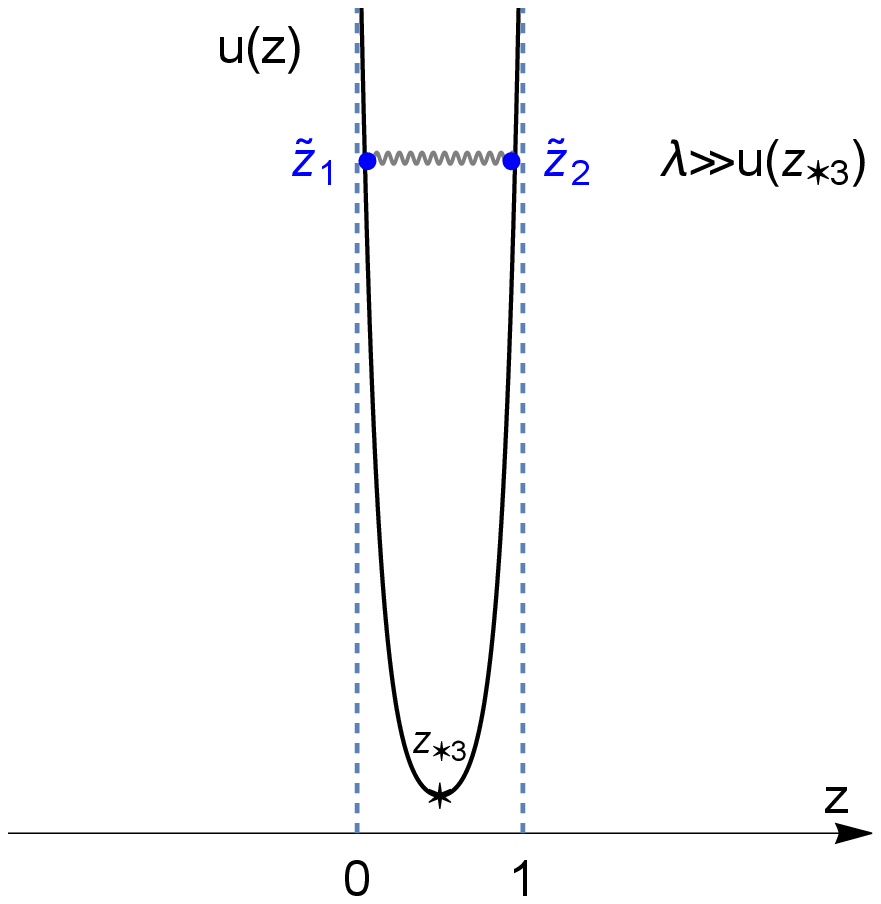}\label{DeepWellPotentialReal}}\hspace{1 cm}
\subfloat[A complex valued singular potential.]{\includegraphics[width=0.45\linewidth]{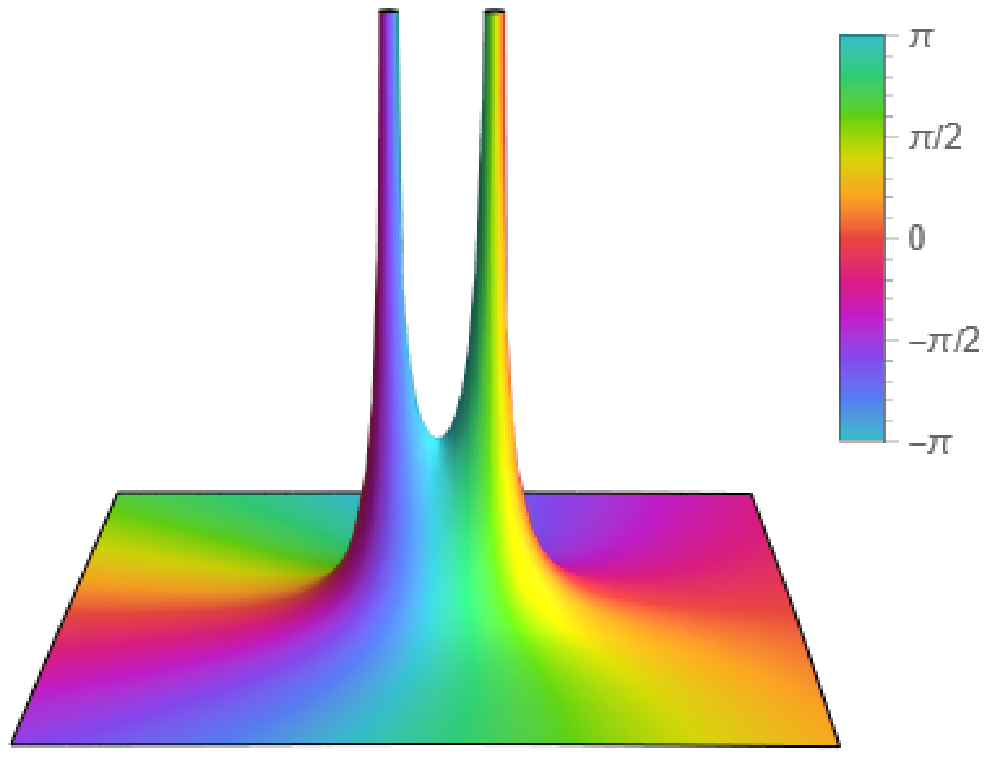}\label{DeepWellPotentialComplex}} \vspace{10 pt}
\caption{The local deep well potential around the saddle point $z_{*3}$. An analogous real valued potential is displayed in Fig. \ref{DeepWellPotentialReal}, the turning points $\widetilde{z}_1\approx 0$ and $\widetilde{z}_2\approx 1$ (that correspond to $\widetilde{x}_1\approx 0$ and $\widetilde{x}_2\approx \mathrm{K}$), so a large energy particle experiences approximately a square well potential of infinite depth with width $L=1$, with energy $\lambda\approx \mu^2/L^2=\mu^2$.
The complex potential is displayed in Fig. \ref{DeepWellPotentialComplex}, 
with colour shade indicating variation of the phase, boundaries of the real potential in Fig. \ref{DeepWellPotentialReal} become singularities of the complex potential at $z=0$ and $z=1$, other two singularities at $z=1/k^2$ and $z\to\infty$ are not included.}
\label{DeepWellPotential}
\end{figure}

To track how the turning points move when the eigenvalue varies, say from the value $\lambda\approx u(x_{*5})$ to a very large value $\lambda\approx\widetilde{\delta}\gg b_s$ according to the relation (\ref{lambdaNear1234th}), one can use the following controlling function,
\begin{equation}
\lambda(\varsigma)=u(x_{*5})+\widetilde{\delta}f(\varsigma),
\end{equation}
to plot the trajectories of turning points, where $\widetilde{\delta}$ is a large quantity and $f(\varsigma)$ smoothly interpolates between $f(0)=0$ and $f(1)=1$. An example of trajectories is presented in Fig. \ref{TrajectoryOfTurningPoints5To1}, it shows that each turning point falls toward a pole. Starting from the configuration of $\lambda\approx u(x_{*5})$, the pattern of falling-toward-a-pole is $z_1\to 1$, $z_2\to 0$, $z_3\to k^{-2}$ and $z_4\to \infty$; the pattern is similar suppose starting from another initial configuration, say from that of $\lambda\approx u(x_{*6})$, it would be $\widehat{z}_1\to 1$, $\widehat{z}_2\to 0$, $\widehat{z}_3\to k^{-2}$ and $\widehat{z}_4\to \infty$.

The turning points associated with the third group of saddle points $z_{*r}$ with $r=1, 2, 3, 4$ have properties different from that of the turning points associated with $z_{*5}$ and $z_{*6}$, here we comment on two aspects.

Firstly, from the formulae (\ref{stationarypoints12}) and (\ref{stationarypoints34}) in Sect. \ref{SaddlePointsOfPotential},
the saddle points are at positions 
\begin{equation}
z_{*1,2}=\sn^2x_{*1,2}\sim\mathcal{O}(k^{-2}),\quad z_{*3,4}=\sn^2x_{*3,4}\sim\mathcal{O}(1).
\end{equation}
One can see that the turning points $\widetilde{z}_{1,2}$ are very close to the saddle points $z_{*3,4}$, $\widetilde{z}_{1,2}-z_{*3,4}\ll\mathcal{O}(1/k^2)$, they cluster around the origin,  as shown in Fig. \ref{TrajectoryOfTurningPoints5To1}. Recall that the eigenvalue at the turning points $\widetilde{z}_{1,2}$ is $\lambda\approx u(x_{*r})+\widetilde{\delta}\sim \mu^2\gg b_s$, and the eigenvalue at the saddle point $z_{*3,4}$ is $\lambda_{*3,4}=u(x_{*3,4})\sim b_s$, so it appears that two close positions are associated with vastly different energies. This happens because near the saddle points $z_{*r}$ with $r =1, 2, 3, 4$ the potential function changes sharply, a small change of the coordinate $z$ could cause a large change of the value of potential $u(z)$.
If we were considering a real valued potential, it can be approximated by the infinitely deep square well potential of unit-width, as in Fig. \ref{DeepWellPotentialReal}, then the energy of a particle in the well is $\lambda\approx \mu^2$ where $\mu$ is an integer.
But we are considering a complex valued potential, the potential has no walls but a few singularities locate in the complex plane, as in Fig. \ref{DeepWellPotentialComplex},
then the quasi-momentum is relaxed to take continuous complex value.
In fact, in the large-$\lambda$ limit the potential can be eliminated using a conformal mapping from $z$-plane back to $x$-plane, except at the four singularities of the potential (\ref{DTVJacobi}), then we have the quantum mechanical problem about a particle of almost free motion, see the argument in the next section around formula (\ref{QuadraticDiffCoulombPotential}).

Secondly, when the value of $\lambda$ is turned to approach the value $\lambda_{*r}=u(x_{*r})$,
or equivalently when the parameter $\widetilde{\delta}$ is turned to vanish, there are two turning points $\widetilde{z}_{t1}$ and $\widetilde{z}_{t2}$ move toward the saddle point $z_{*r}$ and eventually merge with it, similar to that shown in Fig. \ref{TurningAndStationaryPoints}. Taking the first saddle point $z_{*1}$ as the example, one substitutes $\lambda=u(x_{*1})$ into the turning point equation (\ref{TurningPointEq}) to obtain a quadratic polynomial equation of the variable $z=\sin^2x$,
there is a double root with value given by
\begin{equation}
\widetilde{z}_t\big|_{\widetilde{\delta}=0}=(1+\frac{b_1^{1/2}}{b_0^{1/2}})\frac{1}{k^2}-\frac{b_1^{1/2}[(b_0^{1/2}+b_1^{1/2})^2-b_2+b_3]}{2b_0^{1/2}(b_0^{1/2}+b_1^{1/2})^2}+\mathcal{O}(k^2),
\end{equation}
this is the same as the result for the saddle point $z_{*1}=\sn^2x_{*1}$ given by (\ref{stationarypoints12}).
In the process that two turning points approach to and merge with a saddle point,
the parameter $\widetilde{\delta}$ decreases all the way to vanish, the large-$\widetilde{\delta}$ series expansions of turning points are not applicable, therefore, we cannot track the change of expression (\ref{TurningPoints1234th}) with $\widetilde{\delta}$ to see the process of turning points coalescence. The situations for the other three saddle points $z_{*2}$, $z_{*3}$ and $z_{*4}$ are the same.

\section{The quadratic differential associated with the elliptic potential}\label{QuadraticDifferential}

The transformation $z=\sn^2x$ is a conformal mapping from the period parallelogram of potential function in the $x$-plane to the complex $z$-plane;
the equation (\ref{DifferentialEq}) is transformed to the normal form of Heun's differential equation with meromorphic coefficients.
The conformal mapping is associated with a quadratic differential \cite{Strebel1984} which reads
\begin{equation}
\varphi(z)dz^2=\frac{b_0\prod_{t=1}^{4}(z-z_t)}{4z^2(z-1)^2(z-k^{-2})^2}dz^2.
\label{QuadraticDifferential}
\end{equation}
It is related to the leading order WKB integrand 1-form of (\ref{DifferentialFormComplex}) by $v_0(z)dz=\pm\sqrt{\varphi(z)}dz$.
The meromorphic quadratic differential has two kinds of critical points,
the four simple zeros at $z=z_t$ with $t=1, 2, 3 ,4$ are finite critical points,
the four poles of order two at $z=0, 1, k^{-2}$ and $\infty$ are infinite critical points.
One can define a coordinate by the relation $dw=v_0(z)dz=v_0(x)dx$, then a foliation for the complex plane can be introduced,
for a constant $\theta\in[0, 2\pi)$ a family of curves are determined by the condition that $\mbox{arg}(dw)=\theta$ along the curves, in the $w$-coordinate these curves are straight lines.
The length of the curves can be calculated by integrating the invariant line element $\int |dw|=\int |\sqrt{\varphi(z)}dz|$. For a curve containing at least one infinite critical point, the length is infinite. For a curve connecting two finite critical points, or a closed curve, the length is finite.
Finite curves are lifted to the homology cycles of the covering torus surface given by (\ref{SpectralEllipticCurve}).
The quadratic differential and associated foliation have been studied in the context of supersymmetric quantum field theories, the finite length curves are related to BPS states of finite energy \cite{ShapereVafa9910, GMN0907},  the quadratic differential in (\ref{QuadraticDifferential}) is associated with the $SU(2)$ $N_f=4$ theory.
As the potential of differential equation changes continuously, the foliation could undergo discontinuous change of global topology, this phenomenon indicates discontinuity of the spectrum of BPS states with respect to some physical parameters.

Let us zoom in to the regions around a turning point, a pole or a saddle point and investigate the local form of quadratic differential.
Firstly, we consider the general case when the regions around each of these points are moderately separated.
Around a turning point $z_t$, one has
\begin{equation}
\varphi(z)dz^2\approx a_{t}(z-z_t)dz^2,
\end{equation}
where the coefficient $a_{t}=(z-z_t)^{-1}\varphi(z)|_{z=z_t}$, less singular terms are omitted.
This is the quadratic differential for a linear potential;
the local foliation shows Stokes lines of the Airy function. Around a pole $z_p$ of finite value, one has
\begin{equation}
\varphi(z)dz^2\approx a^\prime_{p}(z-z_p)^{-2}dz^2,
\label{QuadraticDiffInverseSquarPotential}
\end{equation}
where the coefficient $a^\prime_{p}=(z-z_p)^{2}\varphi(z)|_{z=z_p}$; and for $z_p=\infty$, one has $\varphi(z)dz^2\approx a^\prime_{\infty}z^{-2} dz^2$.
This is the quadratic differential for a singular inverse square potential;
depending on the complex value of coefficient $a^\prime_{p}$, the local foliation shows ``falling-in'' or ``circling-around" trajectories which are  scale invariant, this is related to the fact that the eigensolution of the linear equation $L\psi(z)=\lambda\psi(z)$ with the operator $L=-\partial_z^2+c/z^2$ has the scaling symmetry $\psi(z)\to \psi(\Lambda z)$, $\lambda\to \Lambda^2 \lambda$.
The region around a saddle points $z_*$ is regular, the foliation is trivial provided zeros and poles locate away from the region.

\begin{figure}[hbt]
\begin{minipage}[t]{0.475\linewidth}
\centering
\includegraphics[width=6 cm]{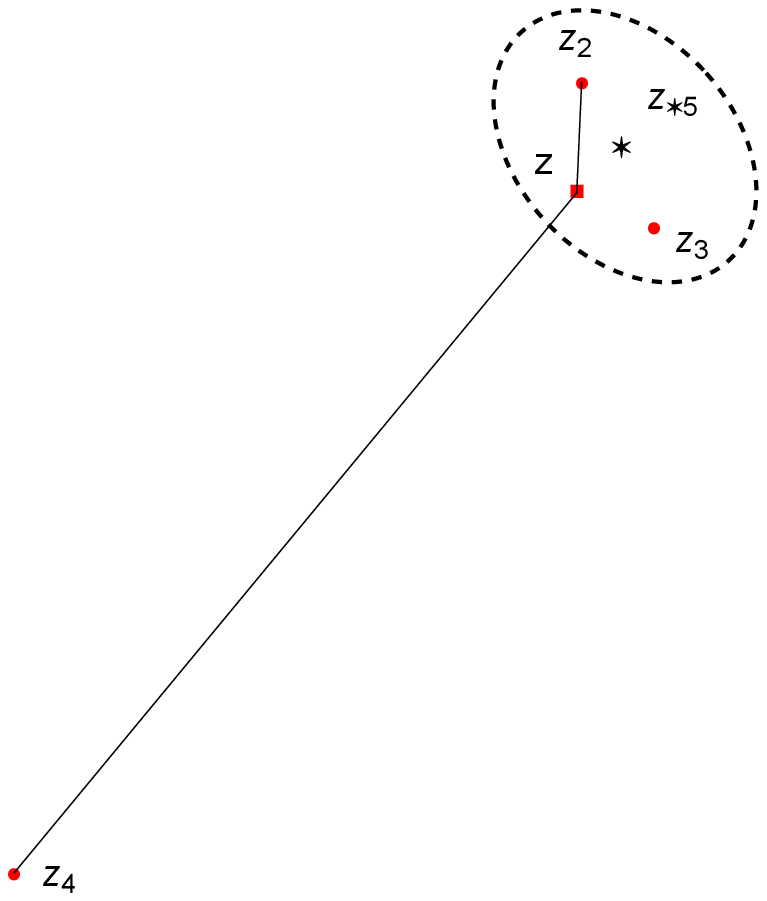}
\vspace{10 pt}
\caption{Since the point $z$ is close to $z_{*5}$, $z_2$ and $z_3$ , away from $z_1$ and $z_4$, then we use $z-z_2=(z-z_{*5})+(z_{*5}-z_2)$, similarly for $z-z_3$, and $z-z_4\approx z_{*5}-z_4$, similarly for $z-z_1$, in (\ref{QuadraticDifferential}) to obtain (\ref{QuadraticDiffHarmonicPotential}).} \label{QuadraDiffApprox5}
\end{minipage}%
\hspace{0.05\linewidth}
\begin{minipage}[t]{0.475\linewidth}
\centering
\includegraphics[width=6 cm]{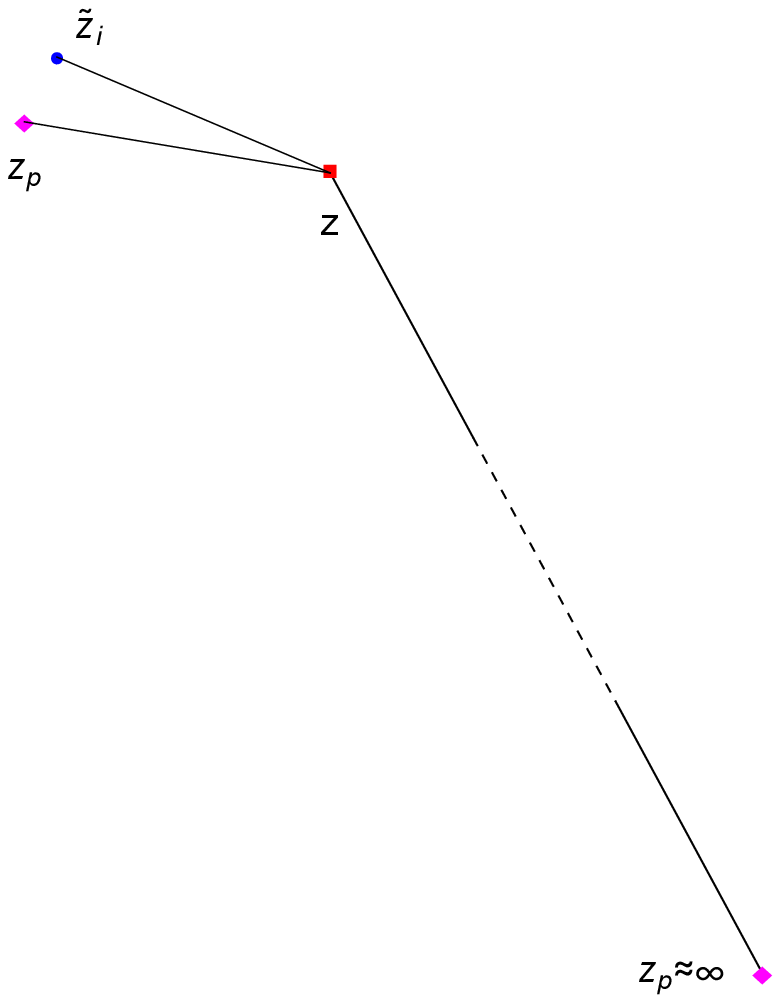}
\vspace{10 pt}
\caption{Since the point $z$ is at finite position, each turning point $\widetilde{z}_t$ is close to a pole $z_p$, in particular $\widetilde{z}_4\approx\widetilde{\delta}/b_0k^2\sim \infty$, then we use $z-\widetilde{z}_t\approx z-z_p$ for finite turning points, and $z-\widetilde{z}_4\approx -\widetilde{z}_4$ in (\ref{QuadraticDifferential}) to obtain (\ref{QuadraticDiffCoulombPotential}).} \label{QuadraDiffApprox1}
\end{minipage}
\end{figure}

Then, we discuss the situation when two or more saddle points, turning points or poles coalesce. One can adjust the value of eigenvalue to the asymptotic region associated to a saddle point $z_{*r}=\sn^2x_{*r}$, that is $\lambda\approx u(x_{*r})\sim \mathcal{O}(bk)$ for $r=5, 6$ or $\lambda\gg b_s\to \infty$ for $r=1, 2, 3, 4$,
the turning points move to the positions given by series solutions in Sect. \ref{TurningPointsOfPotential}.
When the eigenvalue is tuned to $\lambda=u(z_{*5})+\delta$, according to the solution (\ref{TurningPoints5th}),
turning points $z_2$ and $z_3$ move into the region closely around $z_{*5}$ while other two turning points and poles are away from the region.
For a point $z$ in this region, one can apply the approximation to finite distances such as $z-z_1\approx z_{*5}-z_1$, and keep infinitesimal distance such as $z-z_2$ unchanged,
as shown in Fig. \ref{QuadraDiffApprox5},
then the quadratic differential is approximated by
\begin{align}
\varphi(z)dz^2&\approx \frac{b_0(z_{*5}-z_1)(z-z_2)(z-z_3)(z_{*5}-z_4)}{4z_{*5}^2(z_{*5}-1)^2(z_{*5}-k^{-2})^2}dz^2\nonumber\\
&\approx \left[\frac{(b_0-b_1)k^2\delta}{4(b_2-b_3)}-\frac{(b_0-b_1)^{5/2}k^5}{4(b_2-b_3)^{3/2}}(z-z_{*5})^2\right]dz^2.
\label{QuadraticDiffHarmonicPotential}
\end{align}
It corresponds to a harmonic oscillator potential; because $z$ varies in the range spanned by $z_2$ and $z_3$ satisfying (\ref{TuringPoints2and3Distance}), the kinetic term and the potential term have the same order of magnitude.
From the expression (\ref{QuadraticDiffHarmonicPotential}), one can infer the approximate eigenvalue $\delta(\mu)$. 
The standard quantum mechanical equation of harmonic oscillator $-\frac{\hbar^2}{2m}\partial^2_z\psi(z)+\frac{1}{2}m\omega^2z^2\psi(z)=E\psi(z)$ is associated with the quadratic differential $\varphi(z)dz^2=-\frac{1}{\hbar^2}(2mE-m^2\omega^2z^2)dz^2$, 
then the eigenvalue $E=(\mu+\frac{1}{2})\hbar\omega$ implies the leading order dispersion relation $\delta\approx \mp 4i(\mu+\frac{1}{2})(b_0-b_1)^{1/4}(b_2-b_3)^{1/4}k^{1/2}\approx\mp 2\lambda_2^{1/2}(\mu+\frac{1}{2})$. In the local foliation, the two turning points $z_2$ and $z_3$ are connected by a finite curve that passes through the saddle point $z_{*5}$, like a ``tri-atomic molecule". 

For the case when the eigenvalue is tuned to $\lambda=u(z_{*6})+\widehat{\delta}$, the situation is similar.
The quadratic differential is approximated by
\begin{align}
\varphi(z)dz^2&\approx \frac{b_0(z_{*6}-\widehat{z}_1)(z-\widehat{z}_2)(z_{*6}-\widehat{z}_3)(z-\widehat{z}_4)}{4z_{*6}^2(z_{*6}-1)^2(z_{*6}-k^{-2})^2}dz^2\nonumber\\
&\approx \left[\frac{(b_0-b_1)k^2\widehat{\delta}}{4(b_2-b_3)}+\frac{(b_0-b_1)^{5/2}k^5}{4(b_2-b_3)^{3/2}}(z-z_{*6})^2\right]dz^2,
\label{QuadraticDiffHarmonicPotentialAnother}
\end{align}
it implies the leading order dispersion relation $\widehat{\delta}\approx \pm 4(\mu+\frac{1}{2})(b_0-b_1)^{1/4}(b_2-b_3)^{1/4}k^{1/2}\approx \pm 2\widehat{\lambda}_2^{1/2}(\mu+\frac{1}{2})$.

In the other case, when the eigenvalue is tuned to a very large controlling parameter, $\lambda\approx \widetilde{\delta}$, according to the solution (\ref{TurningPoints1234th}), the four turning points each moves very close to a  pole, that is $\widetilde{z}_1\approx 1$, $\widetilde{z}_2\approx 0$, $\widetilde{z}_3\approx k^{-2}$, $\widetilde{z}_4\approx \widetilde{\delta}/b_0k^2\to \infty$, using the approximation shown in Fig. \ref{QuadraDiffApprox1}, the quadratic differential is approximated by
\begin{align}
\varphi(z)dz^2&\approx \frac{-\widetilde{\delta}/4k^2}{z(z-1)(z-k^{-2})}dz^2.
\label{QuadraticDiffCoulombPotential}
\end{align}
It corresponds to four singular Coulomb potentials each locates at a pole.
By the conformal mapping $z=\sn^2x$, the four simple poles in the $z$-plane are transformed to the vertices of the period parallelogram in the $x$-plane,
quadratic differential is transformed to that with no potential in the interior, $\varphi(x)dx^2=\widetilde{\delta}dx^2$, therefore, by the eigenvalue of a free particle one gets the leading order dispersion relation $\widetilde{\delta}\approx \mu^2$. The simple poles at $z=0, 1, k^{-2}$ and $\infty$ are finite critical points, an interpretation is that these simple poles are created from the coalescence of a second order pole and a simple zero \cite{BridgelandSmith1302}. This is the process shown in Fig. \ref{TrajectoryOfTurningPoints5To1} where each turning point moves to merge with a pole.
For example, the turning point at $z=b_2/\widetilde{\delta}+\cdots$ is in close proximity to the pole at $z=0$, forming a ``di-atomic molecule", viewing from a larger distance the foliation around the doublet looks like the foliation of a simple pole at $z\approx 0$.
For the foliation see the figure 28 in the reference \cite{GMN0907}.

We shall point out that all the potentials mentioned above are in one complex space dimension, therefore,
the mathematical treatments of differential equation with these complex potentials are somewhat different from the familiar cases of equation with real potentials in three dimensional space.

It is worth noting that similar foliations appear as the director field surrounding the topological defects of nematic liquid crystal,
in that story point defects are at the place of turning points or poles, line defects are at the place of critical lines,
the shape of the director field is determined by the stability condition of energy \cite{GennesProst1993, Chandrasekhar1992}.

\section{Floquet theorem for ODE with elliptic potential}\label{FloquetTheorem4ODEWithEllipticPotential}

Finally, we explain Floquet theorem for the differential equation with elliptic function potential.
The relevant computations have been presented in previous papers \cite{wh1412, wh1608} and \cite{wh1904},
here we briefly summarise the results, taking into account the corrections caused by the zero point energy for the local potential (\ref{EllipticHillPotentialSN}) and (\ref{EllipticHillPotentialCN}).

The classical Floquet theorem, usually applied for equation with periodic potential $u(x+T)=u(x)$, states that the eigensolutions take the form
\begin{equation}
\psi_\pm(x)=\exp(\pm i\mu x)\phi_\pm(x),
\label{FloquetFunctionReal}
\end{equation}
where $\phi_\pm(x)$ are periodic function, $\phi_\pm(x+T)=\phi_\pm(x)$; the parameter $\mu$ is the Floquet index, it depends on the eigenvalue.
Under translation of coordinate by the period,
\begin{equation}
\psi_\pm(x+T)=\exp(\pm i\mu T)\psi_\pm(x).
\label{FloquetMonodromyReal}
\end{equation}
Therefore, when the eigenfunction is written in the form $\psi(x)=\exp[\int^x v(x^\prime)d x^\prime]$, the index can be computed from integral of $v(x)$ over the period,
\begin{equation}
i\mu=\frac{1}{T}\int_{x_0}^{x_0+T}v(x)dx,
\label{FloquetIndexIntegralReal}
\end{equation}
where the integrand $v(x)$ is one of the two sectors of solution $v_\pm(x)$ for the nonlinear equation $v^2(x)+\partial_x v(x)=u(x)-\lambda$.
From the relation $\mu=\mu(\lambda)$, one can compute the eigenvalue.

For differential equation with elliptic function potential, some new features arises. First of all, an elliptic function potential such as  (\ref{DTVJacobi}) has three periods, $2\mathrm{K}$, $2i\mathrm{K}^\prime$ and the composite $2\mathrm{K}+2i\mathrm{K}^\prime$.
In this case the period $T$, the index $\mu$, the eigenvalue $\lambda$ and the eigenfunctions $\psi_\pm(x)$ all take complex value, so the eigensolutions have rich analytic properties.
Historically, it has been put forward by Hermite the possibility of extending the classical Floquet theorem to ODE with elliptic function potential,
with relations similar to that in (\ref{FloquetFunctionReal}) and (\ref{FloquetMonodromyReal}) for each elliptic period, under the premise that the solution is uniform in the complex plane \cite{Hermite1885};
but it also has been noted that there exist intrinsic difficulties to establish such a theorem due to the singularities of elliptic functions, as commented by Arscott and Wright, etc \cite{Arscott-Wright1969, Sleeman-Smith-Wright1984}.

The results obtained recently in reference \cite{wh1904} provide another perspective on the subject.
According to the explicit asymptotic spectral solutions, to apply the Floquet theorem to elliptic function potential, the eigenfunctions do not always take the form as in (\ref{FloquetFunctionReal}),
instead they acquire the general form
\begin{equation}
\psi_\pm(x)=\exp[i\mu_\pm f(x)]\phi_\pm(x),
\label{FloquetFunctionElliptic}
\end{equation}
where the function $f(x)$ shifts a constant under translation of coordinate by the period,
\begin{equation}
f(x+T)-f(x)=C_T(k^2),
\end{equation}
the constant $C_T(k^2)$ depends on the period \cite{wh1108}.
Then the eigenfunctions are elliptic functions of the second kind, multiplied by constant factor under translation of coordinate by the period, 
$\psi_\pm(x+T)=\exp[i\mu_\pm C_T(k^2)]\psi_\pm(x)$.  
One can regard $\rho=f(x)$ as the canonical coordinate conjugate to the indices $\mu_\pm$, then $C_T(k^2)$ is the period of $\rho$.
The indices $\mu_\pm$ are related to the integral of $v_\pm(x)$ over the corresponding period,
\begin{equation}
i\mu_\pm=\frac{1}{C_T(k^2)}\int_{x_0}^{x_0+T}v_\pm(x)dx.
\label{FloquetExpIntegral}
\end{equation}
For integrals of complex coordinate, one need to choose a contour connecting the base point $x_0$ and its translation image $x_0+T$.
Using the coordinate transformation $z=\sn^2x$, the integral over a period in the $x$-plane is transformed to the integral along a closed contour in the $z$-plane. The period $T$ can be any one of the three elliptic function periods, for each case a particular integral contour must be assigned for (\ref{FloquetExpIntegral}) to obtain the correct result.

Nevertheless, there is an obstacle in applying (\ref{FloquetExpIntegral}) to perform explicit computation.  Owing to the complexity of the potential function, there is no general expression for the integrand $v(x)$ satisfying the equation $v^2(x)+\partial_x v(x)=u(x)-\lambda$.
One thing we can do is to take a step back to study the differential equation with a parametric quantity that is very large or small.
In previous sections, we have showed how to get a large parameter, the two locally expanded form of potential (\ref{EllipticHillPotentialSN}) and (\ref{EllipticHillPotentialCN}) each has a large coupling constant that is not obviously indicated in the differential equation, so that the corresponding integrand can be solved by series expansion.
We also explained that the eigenvalue itself can be used as a large expansion parameter to derive a solution of integrand from the original form of potential (\ref{DTVJacobi}).
By doing this, we obtain three series expansion for the integrand. On the other hand, there are three periods of elliptic function,
$2\mathrm{K}$, $2i\mathrm{K}^\prime$ and $2\mathrm{K}+2i\mathrm{K}^\prime$.
The two facts are closely related: each of the three integrand series solutions is associated with one of the three elliptic function periods,
the three spectral solutions obtained by (\ref{FloquetFunctionElliptic})-(\ref{FloquetExpIntegral})  are Floquet solutions for elliptic function potential. The result obtained in \cite{wh1904, wh1412, wh1608} can be summarised as the following,

\begin{itemize}
\item[(A)] For differential equation with elliptic potential in the local form (\ref{EllipticHillPotentialSN}),
the Floquet theorem applies to the period $T=2i\mathrm{K}^{\,\prime}$ with
\begin{equation}
f(\chi)=\int\frac{d\chi}{\sn\,\chi}=\ln\frac{\dn\,\chi -\cn\,\chi}{k^{\,\prime}\sn\,\chi},\quad C_T(k^2)=i\pi.
\end{equation}
The equation $v^2(\chi)+\partial_\chi v(\chi)=u(\chi)-\delta$ for the integrand leads to the series expansion of $v(\chi)$ for the large coefficient $\lambda_2$,
\begin{equation}
v_\pm(\chi)=\pm\lambda_2^{1/2}\sn\,\chi-\frac{\cn\,\chi\dn\,\chi}{2\sn\,\chi}\pm\frac{1}{\lambda_2^{1/2}}\left(-\frac{3}{8\sn^3\chi}+\frac{1+k^2-4\delta}{8\sn\,\chi}+\cdots\right)+\mathcal{O}\left(\frac{1}{\lambda_2}\right),
\label{IntegrandSNchi}
\end{equation}
the function coefficients of the $\lambda_2$-expansion are Laurent series of $\sn\,\chi$ with singularity at $\chi=0$. Using $\xi=\sn^2\chi$, the integral of $v_\pm(\chi)$ over period $2i\mathrm{K}^{\,\prime}$ in the $\chi$-coordinate is mapped to the integral along a closed contour in the $\xi$-coordinate, the contour encircles the singularity $\xi=0$ and residue theorem can be applied. The resulting series expansion of eigenvalue takes the form,
\begin{equation}
\delta=\mp 2\lambda_2^{1/2}(i\mu_\pm+\frac{1}{2})-\frac{1}{8}(1+k^2)[4(i\mu_\pm+\frac{1}{2})^2+1]+\mathcal{O}\left(\frac{1}{\lambda_2^{1/2}}\right).
\label{EigenvalueSN}
\end{equation}
The formulae are slightly different from those presented in reference \cite{wh1904}. Firstly, the indices appear as $i\mu_\pm$, the definition formulae (\ref{FloquetFunctionElliptic})  (\ref{FloquetExpIntegral}) bring about the imaginary factor,  this is nonessential since the indices are generally complex. 
Secondly, the factor $1/2$ that always accompany indices $\mu_\pm$ comes from integral of the second term in (\ref{IntegrandSNchi}); due to the shift, the indices associated to the same eigenvalue are related by $i\mu_+=-(i\mu_-+1)$. In quantum mechanics, there is a zero point energy term for the  small energy eigenstates around a saddle point of the potential, here the local potential  (\ref{EllipticHillPotentialSN}) is approximately harmonic, $u(\chi)\approx \lambda_2\chi^2$ , the zero point energy term is $1/2$.  This point is missed out in our previous study and is explained in the recent paper \cite{whliu2208}\footnote{Taking into account the $1/2$ factor, one needs to make the following change to the spectral solutions given in reference \cite{wh1904}: for the solution of case (A), $\mu$ is changed to $\mu-i/2$, that means $i\mu$ is shifted to $i\mu+1/2$; for the solution of case (B), $\mu$ is changed to $\mu+k^{\,\prime}/2$, that means $\mu/k^{\,\prime}$ is shifted to $\mu/k^{\,\prime}+1/2$. Notice that in \cite{wh1904} parameters are denoted by letters different from the ones used in this paper.}.

The eigenfunction obtained from the integrand (\ref{IntegrandSNchi})  has an essential singularity at $\sn\,\chi=0$, therefore it is valid in the region away from $\chi=0$. The eigenfunction near $\chi=0$ is represented as a series solution of large $\lambda_2$ with coefficients given by parabolic cylinder functions. The two pieces of eigenfunctions are joined together by connection conditions to form the solution in the region where $|\sqrt{k}\sn\,\chi|<1$. This is the price paid to make the Floquet property apparent.

\item[(B)] For differential equation with elliptic potential in the local form (\ref{EllipticHillPotentialCN}),
the Floquet theorem applies to the period $T=2\mathrm{K}+2i\mathrm{K}^{\,\prime}$ with
\begin{equation}
f(\widehat{\chi})=i\int\frac{d\widehat{\chi}}{\cn\,\widehat{\chi}}=\frac{i}{k^{\,\prime}}\ln\frac{\dn\,\widehat{\chi} + k^{\,\prime}\sn\,\widehat{\chi}}{\cn\,\widehat{\chi}},\quad C_T(k^2)=\frac{\pi}{k^{\,\prime}}.
\end{equation}
The equation $v^2(\widehat{\chi})+\partial_{\widehat{\chi}} v(\widehat{\chi})=u(\widehat{\chi})-\widehat{\delta}$ for the integrand leads to the series expansion of $v(\widehat{\chi})$ for the large coefficient $\widehat{\lambda}_2$,
\begin{equation}
v_\pm(\widehat{\chi})=\pm\widehat{\lambda}_2^{1/2}\cn\,\widehat{\chi}+\frac{\sn\,\widehat{\chi}\dn\,\widehat{\chi}}{2\cn\,\widehat{\chi}}\pm\frac{1}{\widehat{\lambda}_2^{1/2}}\left(-\frac{3k^{\,\prime\,2}}{8\cn^3\widehat{\chi}}+\frac{1-2k^2-4\widehat{\delta}}{8\cn\,\widehat{\chi}}+\cdots\right)+\mathcal{O}\left(\frac{1}{\widehat{\lambda}_2}\right),
\label{IntegrandCNchi}
\end{equation}
the function coefficients of the $\widehat{\lambda}_2$-expansion are Laurent series of $\cn\,\widehat{\chi}$ with singularity at $\widehat{\chi}=\mathrm{K}$. Using $\xi=\sn^2\widehat{\chi}$, the integral of $v_\pm(\widehat{\chi})$ over period $2\mathrm{K}+2i\mathrm{K}^{\,\prime}$ in the $\widehat{\chi}$-coordinate is mapped to the integral along a closed contour in the $\xi$-coordinate, the contour encircles the singularity $\xi=1$ and residue theorem can be applied too. The resulting series expansion of eigenvalue takes the form,
\begin{equation}
\widehat{\delta}=\pm 2\widehat{\lambda}_2^{1/2}k^{\,\prime}(\frac{\mu_\pm}{k^{\,\prime}}+\frac{1}{2})-\frac{1}{8}(1-2k^2)[4(\frac{\mu_\pm}{k^{\,\prime}}+\frac{1}{2})^2+1]+\mathcal{O}\left(\frac{1}{\widehat{\lambda}_2^{1/2}}\right).
\label{EigenvalueCN}
\end{equation}
The factor $1/2$ that accompany indices $\mu_\pm$ comes from integral of the second term in (\ref{IntegrandCNchi}); due to the shift, the indices associated to the same eigenvalue are related by $\mu_+=-(\mu_-+k^{\,\prime})$.
The expression of the first term is consistent with the conclusion from the approximative quadratic differential (\ref{QuadraticDiffHarmonicPotentialAnother}) because $\mu_\pm+k^{\,\prime}/2\approx \mu_\pm+1/2$.
The eigenfunction obtained from the integrand (\ref{IntegrandCNchi}) is valid in the region away from $\widehat{\chi}=\mathrm{K}$.

There is a duality relation between spectral solutions (A) and (B), for example, under
\begin{equation}
k\to \frac{ik}{k^{\,\prime}},\quad \lambda_2\to\frac{\widehat{\lambda}_2}{k^{\,\prime\, 2}},\quad \mu_\pm\to i\left(\frac{\mu_\pm}{k^{\,\prime}}+1\right),\quad \delta\to \frac{\widehat{\delta}}{k^{\,\prime\, 2}},
\end{equation}
the eigenvalue (\ref{EigenvalueSN})  is transformed to (\ref{EigenvalueCN}), here the $1/2$ factor has been taken into account;
the integrand (\ref{IntegrandSNchi}) is transformed to (\ref{IntegrandCNchi}) by another set of relations, the rules are more complicated and can be found in section 4.4.3 of \cite{wh1904}.

\item[(C)] For differential equation with elliptic potential in the original form (\ref{DTVJacobi}),
the Floquet theorem applies to the period $T=2\mathrm{K}$ with
\begin{equation}
f(x)=\int d x=x,\quad C_T(k^2)=2\mathrm{K}.
\end{equation}
The equation $v^2(x)+\partial_x v(x)=u(x)-\lambda$ for the integrand leads to the series expansion of $v(x)$ for the large eigenvalue $\lambda$,
\begin{equation}
v_\pm(x)=\pm i\lambda^{1/2}\pm\frac{u(x)}{2i\lambda^{1/2}}+\frac{\partial_x u(x)}{4\lambda}+\mathcal{O}\left(\frac{1}{\lambda^{3/2}}\right).
\label{Integrandvx}
\end{equation}
Using $\xi=\sn^2x$, the integral of $v_\pm(x)$ over period $2\mathrm{K}$ in the $x$-coordinate is mapped to the integral along a closed contour in the $\xi$-coordinate, the contour encircles a branch cut between $\xi=0$ and $\xi=1$.
In this case, the integration leads to complicated expressions involving complete elliptic integrals $\mathrm{E}(k^2)$ and $\mathrm{K}(k^2)$, see Appendix B in reference \cite{wh1412}. The series expansion of corresponding eigenvalue takes the form,
\begin{equation}
\lambda=\mu_\pm^2+c_0+\mathcal{O}\left(\frac{c_1}{\mu_\pm^2}\right),
\label{EigenvalueDN}
\end{equation}
where the coefficients $c_0$ and $c_1$ are polynomials of $\mathrm{E}(k^2)$, $\mathrm{K}(k^2)$ and $b_s$, they have orders of $c_0\sim\mathcal{O}(b)$, $c_1\sim\mathcal{O}(b^2)$. The indices associated to the same eigenvalue are related by $\mu_+=-\mu_-$.
The eigenfunction obtained from the integrand (\ref{Integrandvx}) is valid in the entire complex plane except at the poles of elliptic functions \cite{wh1904}.
\end{itemize}
Only in the third case (C), the canonical coordinate and its period are the same as that appear in the potential function,
so the classical Floquet theorem applies straightforwardly.
In the cases (A) and (B), one needs to transform to the canonical coordinate so that the eigenfunctions take the standard Floquet form.
In particular, the statement of case (A) can be used to compute a strong coupling spectrum for the Lam\'{e}'s differential equation and the ellipsoidal wave equation, reproducing the same results obtained by other perturbation methods \cite{Muller1966}.

\begin{figure}[htb]
\centering
\subfloat[]{\includegraphics[width=0.378\linewidth]{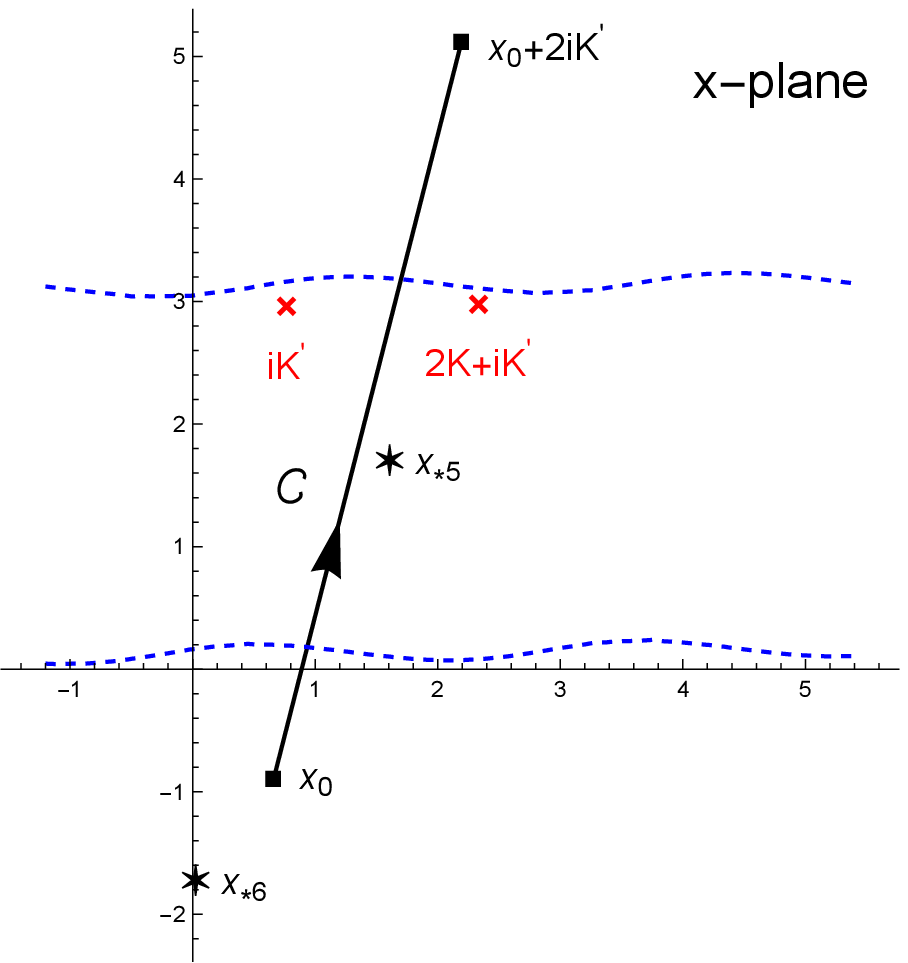}\label{IntegralContourZ5a}}\hspace{2 cm}
\subfloat[]{\includegraphics[width=0.315\linewidth]{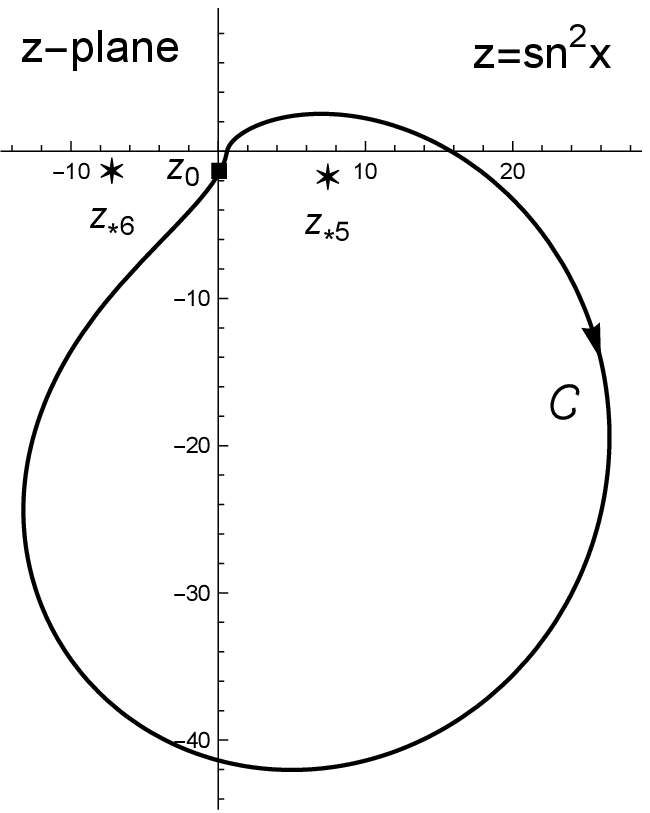}\label{IntegralContourZ5b}} \vspace{10 pt}
\caption{Contour of integrals for the case (A), Fig. \ref{IntegralContourZ5a} plotted in the $x$-plane, Fig. \ref{IntegralContourZ5b} plotted in the $z$-plane, all denoted by $\mathcal{C}$. We have chosen a straight line path in the $x$-plane connecting $x_0$ and $x_0+2i\mathrm{K}^{\,\prime}$ (black squares), the section of line close to the poles $i\mathrm{K}^\prime$ and $2\mathrm{K}+i\mathrm{K}^\prime$ (red cross) in Fig. \ref{IntegralContourZ5a} is mapped to the section of curve stretched downward in Fig. \ref{IntegralContourZ5b}. The path of integral in Fig. \ref{IntegralContourZ5a} should be carefully chosen to ensure the contour in  Fig. \ref{IntegralContourZ5b} does not enclose the other singularity of the integrand $z_{*6}$.
The figures are plotted with $x_0=0.66-0.9i$, and $b_s, k$ with values as previously used in Fig. \ref{DistributionOfPoints5}.}
\label{IntegralContourZ5}
\end{figure}

There is a seemingly contradiction in the above statement, for example, in the case (A), the potential $u(\chi)$ expanded as a Taylor series given by (\ref{EllipticHillPotentialSN}) is convergent in the region where $|\sqrt{k}\sn\,\chi|<1$ as indicated in (\ref{ConvergenceRegionSN}), then what does it mean when doing the integral for the integrand $v(\chi)$ along a path $\mathcal{C}$ extending over the period $2i\mathrm{K}^{\,\prime}$ which certainly exceeds the convergent region, as shown in Fig. \ref{IntegralContourZ5a} ?
In principle, there is  the primary integrand $v(x)$ which is valid everywhere except at the singularities, about which we have no explicit expression;
the integrand $v(\chi)$ results from the local series expansion of $v(x)$ in the region around the saddle point $x_{*5}$.
The integral over period $2i\mathrm{K}^{\,\prime}$ is meant to be applied for $v(x)$.
The contour in the $x$-plane connecting a base point $x_0$ and its periodically translated image $x_0+2i\mathrm{K}^{\,\prime}$ should pass through the nearby of $x_{*5}$; then using $z=\sn^2x$ the path of integral in the $x$-plane is mapped to a closed contour encircling the singular point $z_{*5}$ in the $z$-plane, as shown in Fig. \ref{IntegralContourZ5b}.
The integrand $v(x)$ has a singularity at $x_{*5}$, it becomes the singularity of $v(\chi)$ at $\chi=0$.
By residue theorem, the integral receives contribution only from the singularity; then we can deform the integration contour to contract to a small  region near $z_{*5}$. In this local patch, explicit computation can be carried out using local series expansions of the potential $u(\chi)$ and the integrand $v(\chi)$.
The same argument applies for the local potential $u(\widehat{\chi})$ and integrand $v(\widehat{\chi})$ associated with the region around the saddle point $x_{*6}$.

It should be pointed out that by calculating the spectrum as described, we have not yet achieved the goal of the doubly-period Floquet theorem pursued along the line of Hermite, by which one shall have an explicit expression of quasi-periodic eigenfunction that leads to the above three cases (A), (B) and (C) when applied to the corresponding elliptic period. What we have at the moment are very restricted solutions with apparent Floquet property, namely three asymptotic series expansions of the eigenfunction with a large parameter, two of them are further restricted in some particular regions of the complex coordinate. Hopefully, our study could offer some clue for this classic but obscure subject.

\section*{Acknowledgments}

This work is supported by a grant from CWNU (No. 18Q068).

\end{document}